\documentclass[twocolumn]{aastex631}

\shorttitle{The Tidal Venus Phenomenon}
\shortauthors{Stephen R. Kane \& Emma L. Miles}

\begin{document}

\title{The Tidal Venus Phenomenon: Demographics and Case Studies}

\author[0000-0002-7084-0529]{Stephen R. Kane}
\affiliation{Department of Earth and Planetary Sciences, University of
  California, Riverside, CA 92521, USA}
\email{skane@ucr.edu}

\author[0009-0006-9233-1481]{Emma L. Miles}
\affiliation{Department of Earth and Planetary Sciences, University of
  California, Riverside, CA 92521, USA}


\begin{abstract}

The demographics of terrestrial planets and their orbits reveal a vast
diversity in overall planetary energy budgets. Terrestrial exoplanets
in short-period or eccentric orbits can experience intense tidal
heating that, combined with stellar irradiation, may trigger runaway
greenhouse conditions analogous to Venus. We calculate tidal heating
rates for 143 terrestrial-sized exoplanets with measured
eccentricities and find that, under adopted archive default
eccentricities and constant-$Q$ assumptions, 70\% exceed the extreme
volcanism threshold in tidal flux and 96\% exceed the runaway
greenhouse limit in total zero-albedo flux. We develop a
three-category taxonomy of tidal influence on climate: flux-driven
Venus analogs, tidally dominated planets, and historically compromised
Habitable Zone (HZ) worlds, and apply this framework to five case
studies. TOI-6716~b and TOI-912~b may have exceptionally high tidal
fluxes, potentially serving as examples where tidal dissipation alone
causes them to exceed the runaway greenhouse threshold. TOI-700~d and
LHS~1140~b, though currently below the threshold, were exposed to
above-threshold stellar irradiation during their host stars'
$\sim$0.5--3~Gyr pre-main-sequence phases; whether their atmospheres
survived is testable with JWST. GJ~12~b, already above the threshold
from stellar flux alone, experiences a tidal heat flux of
$\sim$5~W/m$^2$ (comparable to Io) that drives an independent volcanic
pathway to a runaway greenhouse. Three-dimensional climate simulations
show that a temperate atmosphere for GJ~12~b fails to achieve
radiative balance, while a Venus-like CO$_2$-dominated atmosphere
converges to a stable state. We consider observational prospects for
these systems and connections to forthcoming Venus in-situ missions.

\end{abstract}



\section{Introduction}
\label{sec:intro}

The past decade has seen a dramatic expansion in the catalog of known
exoplanets, with confirmed planets now numbering in the thousands
\citep{akeson2013,christiansen2025}. Many of these are
terrestrial-sized planets discovered around a variety of host stars
and occupying a diverse range of system architectures
\citep{ford2014,winn2015}. The wealth of small exoplanets has
intensified interest in the factors that determine their atmospheres
and habitability \citep[e.g.,][and references
  therein]{kane2021e}. Planetary climates are often assumed to depend
primarily on stellar irradiation, including the concept of the
Habitable Zone (HZ), defined by the range of distances yielding
temperate surface conditions for Earth-like planets
\citep{kasting1993a,kane2012a,kopparapu2013a,kopparapu2014,chandler2016,hill2023}. However,
Venus provides a cautionary example of a similarly sized planet that
evolved into a hostile, uninhabitable state despite receiving only
about twice Earth's insolation
\citep{kasting1988c,gillmann2022}. Understanding the divergence of
Venus from Earth is therefore crucial for interpreting exoplanet
observations, particularly runaway greenhouse atmospheres and the
processes that can strip a planet of its water
\citep{angelo2017a,kane2018d,kane2019d,kane2022b,kane2024b,miles2025}. \citet{kane2014e}
defined the ``Venus Zone'' (VZ) around other stars, the region interior
to the HZ where a terrestrial planet is likely to experience a runaway
greenhouse and end up with a Venus-like climate. Since then, numerous
exoplanet discoveries have increasingly populated the VZ with several
hundred terrestrial exoplanets that receive fluxes high enough to
suggest they could be Venus-like rather than Earth-like, even if they
are similar to Earth in size
\citep{ostberg2019,vidaurri2022b,ostberg2023a,kane2026a}.

While high stellar flux is an obvious driver of a runaway greenhouse,
additional internal heat sources can also profoundly affect planetary
energy budgets. In the Solar System, Jupiter's moon Io demonstrates
how tidal gravitational forces can generate substantial heat, as Io's
interior is melted and its surface is peppered with active volcanoes
due to tidal flexing induced by Jupiter and orbital resonances
\citep{peale1979,lainey2009,bierson2021a}. We note that Io is invoked
here as a benchmark for the tidal heat flux that defines the extreme
volcanism regime, not as an atmospheric analog; neither Io nor the
canonical lava world 55~Cnc~e possesses a stable atmosphere, but both
serve as empirical calibration points for the tidal power levels
achievable in real planetary systems. On terrestrial exoplanets, tidal
heating has generally been considered mainly for synchronously
rotating planets (e.g., in very close orbits and/or around M~dwarf
stars) and for its effect on spin states
\citep{jackson2008c,henning2014,driscoll2015,kane2024c}. \citet{barnes2013a}
showed that, for planets orbiting low-mass stars, tidal dissipation
can be high enough to induce complete surface water loss and a runaway
greenhouse scenario. More broadly, the prolonged high-activity phases
of M~dwarf stars can drive atmospheric escape that goes well beyond
desiccation: recent work has demonstrated that nitrogen-dominated
atmospheres on M~dwarf planets can be lost in their entirety under
sustained XUV irradiation \citep{luger2015b,zahnle2017}, a possibility
that must be considered alongside desiccation when assessing the
atmospheric fates of these planets. However, those same tidal effects
can also dampen eccentricity over time, resulting in a temporary
period of extreme tidal heating, followed by a circular orbit in the
HZ with negligible ongoing tides \citep{laskar2012}. Such a planet
might be assumed to be habitable by virtue of its current insolation,
but in reality it would be a sterile, post-runaway greenhouse Venus
analog \citep{barnes2013a,green2019,kane2020e}. This scenario, in
which tidal heating drives a planet into a runaway greenhouse state,
was termed a ``tidal Venus'' by \citet{barnes2013a}. We emphasize that
the term refers to a planet rendered Venus-like by tidal heating, and
not to the tidal forces raised on Venus itself. This further
underscores the need to study Venus as an exoplanet analog to improve
our understanding of the markers of a runaway greenhouse and the
interplay of interior and atmospheric evolution
\citep{kane2019d,way2020}.

Thanks primarily to significant advances in transit and radial
velocity (RV) detection methods, we now have a large sample of
terrestrial exoplanets spanning a range of orbital distances from
their host stars \citep{christiansen2025}. Relatively few of these
have measured eccentricities, since many were discovered via the
transit technique. Those that do are often systems characterized by RV
observations that can provide measurements of the orbit outside of
inferior conjunction \citep{shen2008c,kane2012d,hara2019}. Even so,
the sample of terrestrial planets with known eccentricities is
sufficient to investigate the potential contribution of tidal heating
to the overall energy balance of those planets. One example is the
recently discovered exoplanet Gliese~12~b (hereafter GJ~12~b); a
transiting Earth-sized planet in a 12.76~day orbital period around its
M3 dwarf host star \citep{dholakia2024,kuzuhara2024}. Further RV
observations of the system revealed that the bulk density of the
planet is comparable to Earth, and that the orbit exhibits a
potentially high eccentricity of $e = 0.24 \pm 0.11$
\citep{brady2025b,turner2026a}. Such an eccentricity may have
significant tidal consequences for the planet, including the
accelerated desiccation of surface liquid water, and the transition
into a runaway greenhouse scenario, as described by
\citet{barnes2013a}. GJ~12~b thus provides an opportunity to examine a
concrete example of a potential tidal Venus using actual measured
parameters.

Throughout this paper, the terms ``Earth-like'' and ``Venus-like''
refer specifically to atmospheric compositions: an N$_2$-dominated
atmosphere with trace CO$_2$ in the former case, and a
CO$_2$-dominated atmosphere with Venus-like trace species in the
latter. However, rocky planets around M~dwarfs may experience
formation pathways different from those of the terrestrial Solar
System planets, particularly because giant planets are often absent in
such systems and volatile-rich formation channels may dominate
\citep{raymond2007b,mulders2015b,kane2024a,kane2025a}. The atmospheric
compositions adopted in our climate simulations are therefore treated
as exploratory end states that bracket the expected outcome space,
rather than as predictions from formation models.

In this paper, we investigate the role of tidal heating in pushing
terrestrial exoplanets into runaway greenhouse conditions. We survey
the current terrestrial exoplanet inventory, identifying those that
might be subject to significant tidal energy input, and develop a
taxonomy of tidal influence on planetary climate
states. Section~\ref{sec:sci} details the scientific motivation for
studying tidally induced greenhouse scenarios, including the
implications for terrestrial planet evolution and the connection to
orbital architectures, and introduces a classification framework
distinguishing flux-driven Venus analogs from tidally modified
systems. Section~\ref{sec:demo} presents the demographics of tidal
Venuses through calculations of the tidal heating power and insolation
flux for all confirmed exoplanets within the terrestrial planet
regime. Section~\ref{sec:cases} applies this framework to five
case studies spanning the three taxonomy categories: tidally dominated
planets (TOI-6716~b and TOI-912~b), historically compromised HZ
planets (TOI-700~d and LHS~1140~b), and a tidally compounded Venus
analog (GJ~12~b), including 3D climate simulations for the
latter. Section~\ref{sec:disc} provides a broader discussion of the
prevalence of tidal Venus candidates, their implications for planetary
system architectures and comparative planetology with Venus, and their
relation to JWST and the forthcoming Venus
missions. Finally, Section~\ref{sec:con} summarizes our conclusions
and outlines directions for future work.


\section{Science Motivation}
\label{sec:sci}


\subsection{Venus as a Template for Exoplanet Evolution}

The divergent evolutionary histories of Earth and Venus, despite their
nearly identical masses and bulk compositions, represent a significant
unresolved problem in planetary science and one with profound
implications for exoplanetary science \citep{kane2019d}. Earth has
sustained surface liquid water and biological productivity for
$\sim$4~Gyrs, while Venus has a surface temperature of $\sim$730 K, a
pressure of $\sim$92 bar, and a CO$_2$-dominated atmosphere devoid of
water \citep{kasting1988c}. Whether Venus was once habitable, for how
long, and what triggered its climate catastrophe are central questions
for understanding the frequency of habitable worlds
\citep{way2016,way2020,turbet2021,kane2024b}.
 
Among the plausible triggers for Venus's climate catastrophe, tidal
effects deserve particular attention. \citet{kane2020e} demonstrated
that Jovian planet migration in the early Solar System could have
excited Venus's orbital eccentricity to values as high as $e \sim
0.31$, potentially triggering the runaway greenhouse through a
combination of increased mean insolation, variable perihelion flux,
and tidal dissipation. This work provides a Solar System template for
the kind of dynamical excitation we expect in diverse exoplanetary
systems harboring giant planets \citep{ford2014,winn2015}. The current
near-circular Venusian orbit ($e = 0.007$) could be the endpoint of
tidal circularization following this ancient eccentricity episode,
with the water having been lost during the high-eccentricity phase.
We emphasize that tidal forcing is presented here as one among several
plausible mechanisms, not as an established explanation for Venus's
divergence from Earth. In particular, \citet{turbet2021} demonstrated
that slow rotation alone could have prevented Venus from transitioning
out of the magma-ocean phase, leading to desiccation and CO$_2$
buildup without requiring tidal forcing. The tidal mechanism operates
through a complementary physical channel (energy deposition from
orbital eccentricity rather than rotational climate dynamics) and is
most directly applicable to exoplanetary systems where eccentricity
excitation is common.
 
The study of Venus as an exoplanet is especially important in the era
of space-based exoplanet missions, which have identified a growing
population of sub-2~$R_\oplus$ planets
\citep{dressing2013,petigura2013b,burke2015,dressing2015b,bryson2021}. The
observational bias in utilized detection methods results in many of
these planets receiving insolation fluxes that place them in the VZ of
their stars \citep{ostberg2019,ostberg2023a}. For these planets,
distinguishing between an uninhabitable Venus-like state and a
potentially habitable Earth-like state is a primary goal of JWST
atmospheric characterization programs
\citep{ehrenreich2012a,barstow2016a,ostberg2023c}. The tidal Venus
mechanism adds a new dimension to this problem, since even planets
outside the VZ may have evolved toward a Venus-like state if they have
experienced sufficient tidal heating \citep{barnes2013a}.


\subsection{The Role of System Architecture in Terrestrial Planet Evolution}

A growing body of evidence suggests that system architecture plays a
decisive role in terrestrial planet evolution
\citep{morbidelli2012a,nesvorny2018c,zhu2021}. Gravitational
interactions between planets can excite and maintain non-zero
eccentricities on inner terrestrial planets
\citep{barnes2013a,kane2024c}. In our Solar System, Jupiter's current
position at 5.2~AU keeps Earth's eccentricity oscillating between 0
and $\sim$0.06 on timescales of $\sim$400 kyr \citep{laskar1988b}. For
Venus, this modulation is smaller ($e \lesssim 0.07$) but the
situation was potentially much more dramatic during Jupiter's early
migration \citep{kane2020e}.

In exoplanetary systems, the diversity of giant planet architectures
is far greater than in our solar system. Many systems discovered by RV
surveys contain giant planets with eccentricities lying within the
full range permissible by bound orbits
\citep{winn2015,fulton2021,rosenthal2022}, and some harbor multiple
giant planets on mutually excited orbits. For example, the HD~104067
system harbors two giant planets in eccentric orbits that excite the
inner terrestrial planet candidate to eccentricities that produce
tidal power comparable to the stellar irradiation
\citep{kane2024c}. The LP~791-18 system offers another example, where
dynamical interactions between a sub-Neptune and a super-Earth force
the outer, temperate planet into an eccentric orbit with substantial
tidal heating \citep{peterson2023}. For single-planet systems, or
systems where outer planets have not yet been detected, the story is
incomplete and may be the result of past planet-planet interactions
\citep{kane2014b,kane2016d,carrera2019b}.


\subsection{The Tidal Venus Threshold}

The idea that tidal forces can profoundly influence a planet's
habitability broadens the traditional view that stellar insolation
alone can dominate a planetary atmospheric energy budget. For planets
around low-mass stars (late K and M dwarfs), there are several reasons
to suspect tidal heating could be important. First, the HZs of these
stars lie very close-in (e.g., $<$0.1--0.3~AU for M dwarfs of
0.1--0.3~$M_\odot$), meaning a potentially habitable planet would be
in a short-period orbit where tidal effects are stronger. Second, many
M dwarfs are observed to have multi-planet systems with compact
architectures, in which planet-planet perturbations or resonances can
sustain nonzero eccentricities
\citep{fabrycky2012,kane2012d,hadden2014,vaneylen2015}. Tidal
dissipation within the planet can then act as a long-term energy
source for the atmosphere and surface, potentially impacting long-term
habitability prospects \citep{barnes2009b,bolmont2026}.

The critical question for the tidal Venus phenomenon is at what energy
flux a runaway greenhouse is triggered. For stellar irradiation alone,
this threshold has been estimated using 1D radiative-convective models
as $\sim$1.06 $S_\oplus$ for a planet identical to Earth
\citep{kopparapu2013a,kopparapu2014}, though 3D general circulation
models (GCMs) suggest the threshold may be higher, closer to $\sim$1.4
$S_\oplus$ for slowly rotating planets
\citep{leconte2013c,yang2014b}. For the purpose of this paper, we
adopt the \citet{kopparapu2013a,kopparapu2014} runaway greenhouse flux
of $S_{\rm RG} = 1.06$~$S_\oplus$ as a conservative threshold for the
onset of the runaway, consistent with the approach of
\citet{barnes2013a}. \citet{barnes2013a} calculated that an Earth-mass
planet around a 0.2~$M_\odot$ star at the inner edge of the HZ could
experience tidal power on the order of $10^{15}$--$10^{16}$~W under
certain eccentricity conditions, enough to trigger water loss on
$\sim$$10^8$~year timescales. For comparison, the tidal heating of Io
results in a total tidal dissipation of $10^{14}$~W
\citep{bierson2021a}. If the stellar irradiation alone places a planet
below the runaway greenhouse threshold, but tidal heating pushes the
effective total flux above $S_{\rm RG}$, the planet may still undergo
a tidal greenhouse. Furthermore, because tidal heating decreases as
the orbit circularizes, a planet may have experienced a ``Tidal
Venus'' episode in the past (when eccentricity was higher) and now
reside on a circular orbit within the HZ; permanently desiccated but
no longer tidally heated.


\subsection{A Classification Framework for Tidal Influence}
\label{sec:taxonomy}

In the context of the tidal Venus phenomenon, it is important to
distinguish several related but physically distinct categories of
planetary climate influence. We adopt the following classification
framework, which we apply to the case studies in
Section~\ref{sec:cases}.

The first category consists of flux-driven Venus analogs:
planets whose stellar irradiation alone places them above the runaway
greenhouse threshold, regardless of tidal heating. These planets
reside in the Venus Zone \citep{kane2014e} and would be classified as
Venus analogs on the basis of their orbital location and stellar flux
alone \citep{ostberg2023a,mcintyre2023a}. Tidal heating may compound
the Venus-like state through Super-Io volcanism and outgassing but is
not the decisive factor in the classification \citep{mahapatra2023}.
When such tidal compounding is present, these planets may also be
referred to as ``tidally compounded Venus analogs.''

The second category comprises tidally dominated planets: planets whose
tidal heating flux alone exceeds the runaway greenhouse threshold of
295~W/m$^2$, meaning that tidal dissipation is the primary energy
source driving the planet's climate state. These planets are above the
runaway greenhouse threshold regardless of their stellar
irradiation. Several planets in the current sample occupy this regime,
though their eccentricities require confirmation through further RV
observations \citep{barnes2013a,foley2018a}.

The third category encompasses historically compromised HZ planets:
planets that are currently below the runaway greenhouse threshold from
both stellar and tidal flux but that were exposed to above-threshold
conditions during the host star's pre-main-sequence (pre-MS)
contraction phase. For M~dwarf hosts, this phase lasts
$\sim$0.5--3~Gyr \citep{ramirez2014c,baraffe2015}, during which the
stellar luminosity can exceed the present-day value by factors of
several. Any eccentricity present during this epoch, which for long
circularization timescales persists to the present day, would have
contributed tidal heating atop the already elevated stellar flux.
Whether these planets retained their atmospheres depends on their
mass, volatile inventory, and the duration of the above-threshold
epoch \citep{luger2015b,zahnle2017,mcintyre2022}.

This classification extends the existing Venus Zone framework
\citep{kane2014e} by incorporating tidal heating and pre-MS stellar
evolution as additional dimensions that determine the evolutionary
fate of terrestrial exoplanets around low-mass stars.


\section{Demographics of Tidal Venuses}
\label{sec:demo}

Here, we examine the stellar and tidal energy components for known
terrestrial exoplanets in the context of various transitional limits.


\subsection{Stellar and Tidal Energy}
\label{sec:energy}

To evaluate the potential for tidal greenhouse conditions across the
exoplanet population, we calculated two key energy quantities for each
relevant planet: (1) the incident stellar flux absorbed by the planet,
and (2) the tidal power dissipated within the planet due to
eccentricity tides. The stellar irradiance received by the planet,
$F_p$, was computed as:
\begin{equation}
  F_p = \frac{L_\star}{4 \pi a^2}
  \label{eq:flux}
\end{equation}
where $L_\star$ is the stellar luminosity and $a$ is the semi-major
axis of the planet. The effective flux absorbed by the planet is often
approximated by considering the cross-sectional area of the planet
divided by the surface area, leading to an absorbed component of
$F_{\rm star} = F_p / 4$. For example, these values for Earth at its
semi-major axis are $F_p \approx 1361$~W/m$^2$ and $F_{\rm star}
\approx 340$~W/m$^2$ \citep{kopp2011a,kopp2023}. Note that flux
absorbed at the surface is dependent on the Bond albedo, $A$, such
that the absorption scales with $(1-A)$. For now, we assume $A = 0$ to
compute a maximal absorbed power, effectively treating the planet as
an ideal blackbody. Under this convention, Earth itself has $F_{\rm
  star} \approx 340$~W/m$^2$, which exceeds the runaway greenhouse
threshold of 295~W/m$^2$ (Section~\ref{sec:limits}); it is Earth's
albedo of $\sim$0.3 that reduces the absorbed flux below the
threshold and maintains temperate conditions. This zero-albedo
framework provides a conservative upper bound on the planetary energy
budget, with the role of albedo explored separately for individual
systems (Section~\ref{sec:energy_budget}).

The tidal heating power $P_{\rm tide}$ was computed using the formula
for a homogeneous, viscoelastic body in a fixed eccentric
orbit, expressed as:
\begin{equation}
  P_{\rm tide} = \frac{21}{2} \frac{k_2}{Q} \frac{G^{3/2}
    M_\star^{5/2} R_p^5 e^2}{a^{15/2}}
  \label{eq:tides}
\end{equation}
where $G$ is the gravitational constant, $M_\star$ is the mass of the
host star, $R_p$ is the radius of the planet, and $e$ is the orbital
eccentricity \citep{squyres1983b,meyer2007}. The parameter $k_2$ is
the planet's second-order Love number and $Q$ is the tidal quality
factor (a dimensionless measure of how dissipative the planet is). We
adopt conservative values of $k_2 = 0.3$ and $Q = 50$ for terrestrial
planets, consistent with Earth and Venus models
\citep{zhang1992c,henning2009}. We emphasize that $P_{\rm tide}$
scales extremely steeply with orbital distance: $P_{\rm tide} \propto
a^{-15/2}$ for fixed $M_\star$. This extreme distance dependence means
that only relatively close-in planets will experience significant
tidal power unless $e$ is very large to compensate at wider
separations. To convert the tidal power to a flux at the surface, we
simply divide by the surface area of the planet: $F_{\rm tide} =
P_{\rm tide} / (4 \pi R_p^2)$.

This formalism treats the planet as a homogeneous body and does not
resolve where within the interior tidal dissipation is deposited. In
reality, the depth and mechanism of dissipation depend on interior
structure, mantle viscosity, and rheology, and can significantly
affect the efficiency of eccentricity damping and the coupling between
tidal heating and surface volcanism
\citep{henning2009,renaud2018}. Advanced rheological models (e.g.,
Andrade or Sundberg-Cooper rheologies) can produce tidal heating rates
differing by an order of magnitude from the constant-$Q$ Maxwell
approximation for warm rocky interiors
\citep{renaud2018}. Furthermore, multilayer tidal friction
calculations show that the homogeneous approximation is generally
adequate for purely rocky planets, but may break down for bodies with
icy layers \citep{bolmont2020b}. The constant-$Q$ approach adopted
here is standard in the exoplanet tidal heating literature
\citep{barnes2013a,heller2015b,driscoll2015,kane2024c} and is
appropriate for the demographic scope of this work, given that
interior structures are unconstrained for the planets in our
sample. Refinement of these estimates using layered interior models is
an important direction for future work as interior constraints become
available.

We further note that Equation~\ref{eq:tides} captures only the
eccentricity (radial) component of the tidal response, which is the
dominant contribution for the eccentric orbits that are the focus of
this work. Additional tidal components include obliquity tides arising
from spin-orbit misalignment \citep{heller2011a} and, for
non-synchronously rotating planets, thermal or rotational
tides. Obliquity tides can rival the eccentricity contribution when
the planetary obliquity is large, but tidal processes tend to erode
obliquity over timescales shorter than the age of the system for
close-in planets \citep{heller2011a}. Since the spin states and
obliquities of the planets in our sample are unconstrained by current
observations, we adopt the eccentricity tide as the representative
tidal component and treat our tidal fluxes as lower bounds in cases
where significant obliquity may persist. We also note that our
solid-body constant-$Q$ formulation omits dissipation in oceans and
atmospheres. For ocean-bearing worlds, such as present Earth, fluid
tidal dissipation can exceed the solid-body value by more than an
order of magnitude \citep{bolmont2020b}.

Finally, we define the total heating flux $F_{\rm total} = F_{\rm
  star} + F_{\rm tide}$. These values, combined and independently, are
used to approximate the potential state of the planet in relation to
various tipping points that can influence the interior, surface, and
climate evolution.


\subsection{Energy Limits for Climate/Surface Transitions}
\label{sec:limits}

The combined stellar and tidal energy flux experienced by a planet can
drive it across critical thresholds that profoundly alter its surface
environment, geological activity, and atmospheric state. Here we
identify three such limits that are directly relevant to assessing the
tidal Venus phenomenon across the known terrestrial exoplanet
population. Two of the three limits (the runaway greenhouse and molten
surface thresholds) depend on the total flux, $F_{\rm total}$, because
both the stellar and tidal contributions ultimately feed the
atmospheric and surface energy budget. The extreme volcanism
threshold, however, depends only on the tidal flux, since it
characterizes a geophysical state of the planetary interior driven by
tidal dissipation independently of the stellar irradiation
environment.


\subsubsection{Extreme Volcanism Limit}

The extreme volcanism limit characterizes the transition from levels
of tectonic and volcanic activity similar to present Earth to a state
of pervasive, globally dominant volcanism analogous to Jupiter's moon
Io. Io experiences a mean surface heat flux of $\sim$2~W/m$^2$ from
tidal dissipation driven by its orbital resonance with Europa and
Ganymede \citep{lainey2009,bierson2021a}. \citet{heller2013a} and
\citet{driscoll2015} adopt this value as the lower boundary of the
``Super-Io'' regime, where tidal energy dominates the internal heat
budget and drives eruption rates orders of magnitude greater than
present-day Earth. We therefore adopt a tidal flux of:
\begin{equation}
  F_{\rm tide} \geq 2~\mathrm{W/m}^2
  \label{eq:ev_limit}
\end{equation}
as the extreme volcanism limit. Because this threshold is set by the
internal dissipation rate of the planet, it depends only on $F_{\rm
  tide}$, not on the stellar flux.

Crucially, although crossing this threshold does not directly trigger
a runaway greenhouse through thermal forcing alone, it has an indirect
pathway to driving such a climate transition through geochemical
forcing. A Super-Io planet releases enormous quantities of CO$_2$ and
SO$_2$ through volcanism on timescales potentially shorter than
atmospheric sequestration, significantly amplifying the greenhouse
effect and potentially pushing an already warm planet past the moist
greenhouse threshold \citep{noack2017a}. This volcanic outgassing
pathway is especially relevant for planets already near the runaway
greenhouse boundary from stellar irradiation alone.


\subsubsection{Runaway Greenhouse Limit}

The runaway greenhouse occurs when the outgoing longwave radiation
(OLR) from a moist atmosphere saturates at a critical flux $F_{\rm
  crit}$, above which the atmosphere can no longer maintain radiative
equilibrium and all surface water evaporates. This saturation arises
because a water-dominated atmosphere becomes optically thick in the
thermal infrared once it is warm enough, fixing the emission
temperature at the upper troposphere regardless of surface
temperature. The resulting OLR limit has been previously calculated
using 1D radiative-convective models \citep{nakajima1992,kasting1988c}
and yields values in the range 280--310~W/m$^2$ for an Earth-mass
planet \citep{goldblatt2013,barnes2013a}. Within this range,
\citet{goldblatt2013} obtained a value of $\sim$282~W/m$^2$ using a
line-by-line radiative transfer calculation, which has been broadly
confirmed by subsequent studies \citep{schaefer2016,lichtenberg2021b}.
The higher end of the range reflects differences in opacity databases,
treatment of continuum absorption, and assumed atmospheric
composition. Following \citet{heller2015b}, we adopt:
\begin{equation}
  F_{\rm total} = F_{\rm star} + F_{\rm tide} \geq F_{\rm crit} =
  295~\mathrm{W/m}^2
  \label{eq:rg_limit}
\end{equation}
as the runaway greenhouse limit for an Earth-mass planet. This value
lies at the midpoint of the classical range and is consistent with the
approach of \citet{barnes2013a}. Note that adopting a lower value,
such as the \citet{goldblatt2013} estimate of 282~W/m$^2$, would shift
the runaway greenhouse curve in Figure~\ref{fig:exo} marginally inward
but would not affect any of our qualitative conclusions, since
GJ~12~b's stellar flux of $\sim$573~W/m$^2$ already exceeds even the
highest published OLR limit by nearly a factor of two. The choice of
$F_{\rm crit}$ within the published range is therefore not
consequential for the specific case of GJ~12~b, though it could matter
for planets closer to the threshold. Because tidal dissipation
deposits heat directly into the planetary interior and must be
radiated regardless of albedo, its contribution to $F_{\rm total}$ is
energetically equivalent to absorbed stellar flux from the perspective
of the atmospheric energy budget. The combined nature of this limit is
the defining feature of the tidal Venus scenario, whereby a planet
that would be subcritical from stellar flux alone can cross the
threshold when tidal heating is accounted for.


\subsubsection{Molten Surface Limit}

At sufficiently high total flux, the planet's surface itself becomes
molten. The condition for a sustained liquid silicate surface is that
the total incoming energy exceeds the thermal emission of a surface at
the silicate solidus temperature, $T_{\rm melt} \approx 1600$~K
\citep{driscoll2015}. Treating the surface as an ideal blackbody at
zero albedo (an appropriate approximation since liquid rock has a Bond
albedo of only $\sim$0.05--0.1) the Stefan-Boltzmann condition yields:
\begin{equation}
  F_{\rm total} = F_{\rm star} + F_{\rm tide} \geq \sigma T_{\rm
    melt}^4 \approx 3.7 \times 10^5~\mathrm{W/m}^2,
  \label{eq:ms_limit}
\end{equation}
or equivalently $\sim$370~kW/m$^2$. This threshold has been applied
to tidally heated exoplanets in the context of lava worlds and super-Io
bodies \citep{driscoll2015,kane2024c}. A planet exceeding this limit
would resemble a permanently molten, geologically resurfacing world,
incapable of sustaining any stable surface chemistry. Like the runaway
greenhouse limit, this threshold depends on the total flux budget and
therefore reflects the combined contribution of both stellar and tidal
heating. In practice, only the most extremely irradiated and/or
tidally heated planets in the known population reach this regime, as
discussed in Section~\ref{sec:dist}.


\subsection{Energy Budget Distribution for Small Planets}
\label{sec:dist}

\begin{figure}
  \includegraphics[width=\linewidth]{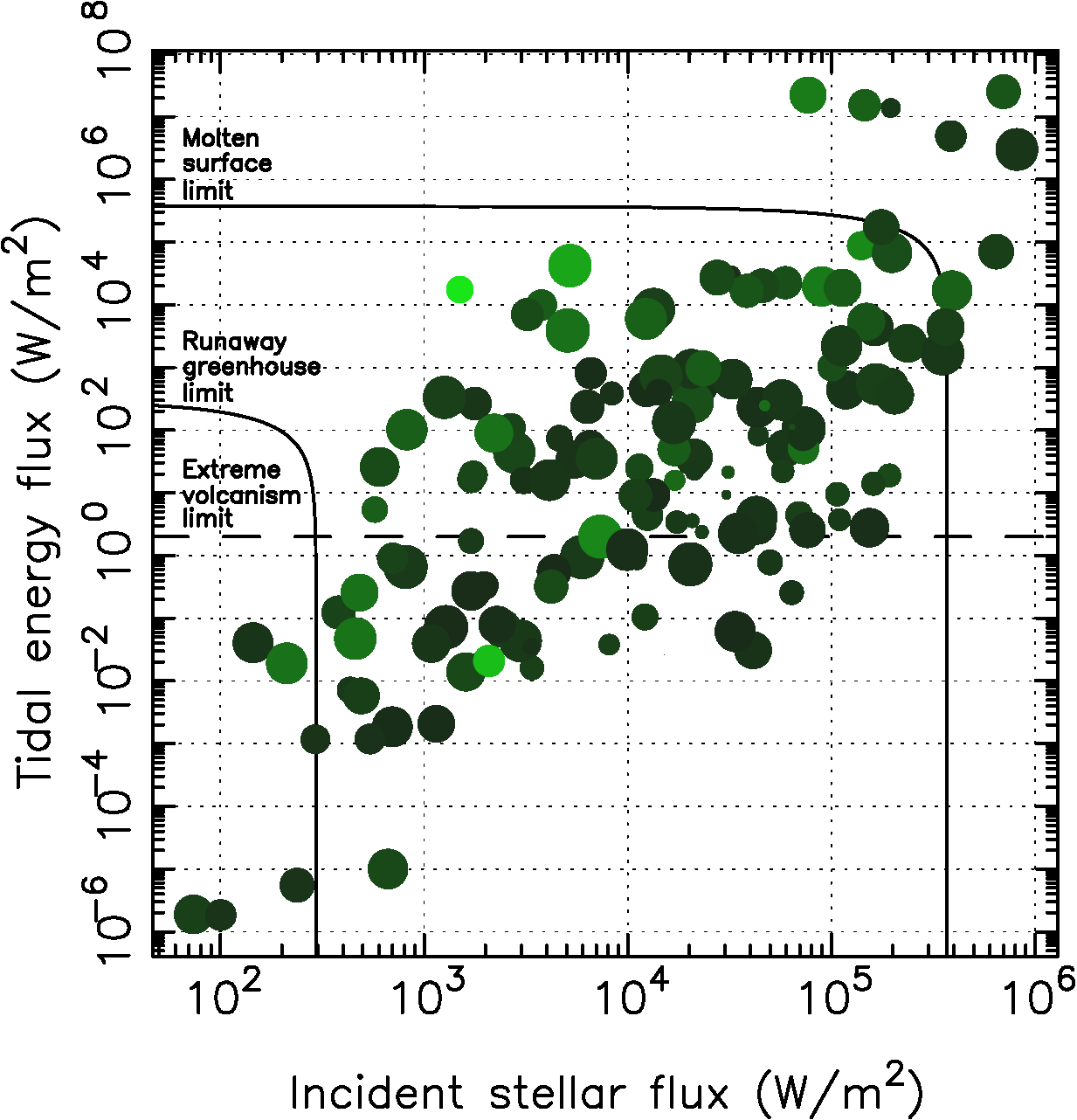}
  \caption{Tidal energy flux versus incident stellar flux (averaged
    over the planetary surface) for known terrestrial exoplanets ($R_p
    \le 2$~$R_\oplus$). The sizes of the data points are
    logarithmically scaled to the planetary radius. The colors are
    linearly scaled to the orbital eccentricity, where dark green is
    low eccentricity and bright green is high eccentricity. The
    horizontal dashed line marks the extreme volcanism energy limit
    ($F_{\rm tide} = 2$~W/m$^2$), and the two solid curves indicate
    the runaway greenhouse limit ($F_{\rm total} = 295$~W/m$^2$) and
    molten surface limit ($F_{\rm total} = 3.7\times10^5$~W/m$^2$).
    Planets lying above and to the right of each curve exceed the
    corresponding threshold.}
  \label{fig:exo}
\end{figure}

To investigate the demographics of potential tidal Venus scenarios, we
focus on planets with radii $R_p < 2~R_{\oplus}$, which are likely to
be rocky \citep{rogers2015a} and thus comparable to Earth and Venus in
composition. For consistency, we used the default planet parameters
compiled in the NASA Exoplanet Archive \citep{christiansen2025}, which
were extracted on 2026 April 3. We include only those planets with the
necessary parameters to perform the calculations described in
Section~\ref{sec:energy} and whose eccentricity values are greater
than zero, yielding a sample of 143 planets. Note that the default
eccentricity values reported by the archive are heterogeneous in
quality: while many are well-constrained detections from RV orbital
fits, others represent upper limits or poorly constrained posterior
modes with uncertainties spanning the full range to zero (e.g., $e =
0.88^{+0.0}_{-0.88}$ for TOI-6716~b; \citealt{scott2026}). The tidal
fluxes computed for such planets should be regarded as upper bounds
pending further RV follow-up.

The calculated fluxes are presented in Table~\ref{tab:demo}, sorted by
decreasing stellar flux, and are shown in Figure~\ref{fig:exo}. The
data show that the overwhelming majority (137 out of 143 planets, or
96\%) have total fluxes $F_{\rm total} \geq 295$~W/m$^2$, placing them
above the runaway greenhouse limit under the zero-albedo convention
adopted in Section~\ref{sec:energy}. A nonzero Bond albedo would
reduce the absorbed stellar flux by a factor of $(1-A)$, potentially
moving some planets below the threshold; the role of albedo is
explored for the GJ~12~b case study in Section~\ref{sec:energy_budget}. This
high fraction primarily reflects the strong
observational bias in the current exoplanet census toward short-period
planets around all stellar types, most of which are far too irradiated
to be habitable regardless of tidal contributions
\citep{winn2015,bryson2020b}. Only six planets in the sample fall
below the runaway greenhouse threshold: Kepler-441~b, Kepler-186~f,
LHS~1140~b, Kepler-296~f, Kepler-442~b, and TOI-700~d. These are all
confirmed or candidate HZ worlds with minimal eccentricities and
negligible tidal heating, and their presence in the lower-left of
Figure~\ref{fig:exo} confirms the self-consistency of the flux
calculations.

Eight planets exceed the molten surface limit ($F_{\rm total} \geq
3.7\times10^5$~W/m$^2$): K2-211~b, K2-147~b, TOI-1238~b, GJ~367~b,
TOI-500~b, 55~Cnc~e, Kepler-323~b, and KOI-94~b. GJ~367~b is a
sub-Earth-sized ultra-short-period planet ($P = 0.322$~days) in a
highly eccentric orbit, and its enormous tidal flux
($\sim$$1.4\times10^7$~W/m$^2$) dominates its total energy budget,
despite also receiving intense stellar irradiation. Similarly,
K2-211~b and K2-147~b achieve their extreme tidal fluxes
($\sim$$2.5\times10^7$ and $2.2\times10^7$~W/m$^2$, respectively)
through a combination of close orbital distances and relatively high
eccentricities. TOI-1238~b is another extreme case with tidal flux
exceeding $10^7$~W/m$^2$, classifying it among the most tidally active
terrestrial planets known. For comparison, 55~Cnc~e reaches the molten
surface regime primarily through stellar flux
($8.1\times10^5$~W/m$^2$) rather than tidal heating, and is a
canonical lava world for which high-precision thermal emission
measurements have been obtained \citep{demory2016b}.

The horizontal dashed line at 2~W/m$^2$ marks the extreme volcanism
limit. A substantial fraction of the sample (100 out of 143 planets,
or 70\%) have tidal fluxes exceeding this threshold. This reflects the
concentration of the sample at short orbital periods where tidal
effects are amplified. Many of these planets are so highly irradiated
that the tidal contribution is negligible by comparison to their total
energy budget. However, as discussed in Section~\ref{sec:limits}, the
volcanic outgassing pathway operates independently of whether the
stellar flux alone crosses the runaway greenhouse threshold, meaning
Super-Io volcanism can push a geochemically susceptible planet over
the atmospheric tipping point even when the tidal flux is subdominant
in the total energy budget \citep{noack2017a}.

A distinct population of planets exhibits tidal fluxes that are
comparable to or exceed their stellar irradiation, placing them in the
tidally dominated regime defined in Section~\ref{sec:taxonomy}. For
example, TOI-912~b has a stellar flux of
$5.2\times10^3$~W/m$^2$, but a tidal flux of
$4.2\times10^4$~W/m$^2$ produced entirely from its orbital
eccentricity, meaning that tidal dissipation dominates the total
energy budget by a factor of $\sim$8.
TOI-6716~b, LHS~1678~b, TOI-1749~b, and TOI-2096~b
similarly have their tidal fluxes comparable to or exceeding their
stellar irradiation, making them among the most compelling tidal Venus
candidates in the current sample. However, as noted above, several of
these planets have eccentricities that are upper limits rather than
firm detections, and their extreme tidal fluxes are contingent on
confirmation of their orbital parameters (see
Section~\ref{sec:tidal_dom}).

Separately, GJ~12~b illustrates the tidally compounded regime. Its
stellar flux (surface averaged)
of $\sim$573~W/m$^2$ already places it above the runaway greenhouse
limit from irradiation alone. Its tidal flux of $\sim$5.4~W/m$^2$ is
non-negligible, given that it exceeds the extreme volcanism threshold,
classifying GJ~12~b as a candidate Super-Io volcano world. This tidal
flux is too small to drive a runaway greenhouse directly, but the
volcanic outgassing it drives represents an independent, geochemically
mediated pathway to a runaway state, as discussed in detail in
Section~\ref{sec:gj12b}.

For Solar System context, present-day Earth and Venus both fall
outside the plotted region of Figure~\ref{fig:exo}. In terms of
stellar flux, Earth ($F_{\rm star} \approx 340$~W/m$^2$) and Venus
($F_{\rm star} \approx 650$~W/m$^2$) both lie above the runaway
greenhouse threshold under the zero-albedo convention, with Venus's
higher insolation reflecting its closer orbit. Their eccentricity
tidal fluxes, however, are negligible: adopting their current
eccentricities ($e_\oplus = 0.0167$, $e_{\rm V} = 0.0068$), the
solid-body eccentricity tidal flux is $\sim$$10^{-9}$~W/m$^2$ for
both, roughly nine orders of magnitude below the extreme volcanism
threshold and far below the lower boundary of
Figure~\ref{fig:exo}. This is consistent with the well-established
result that tidal heating is dynamically insignificant for the present
Solar System terrestrial planets \citep{bolmont2020b}. For Earth, the
total tidal dissipation is dominated not by the solid-body response
but by ocean tides, which contribute $\sim$3.7~TW globally
\citep{bolmont2020b}, equivalent to a surface flux of
$\sim$$7\times10^{-3}$~W/m$^2$, which also is well below the extreme
volcanism threshold.

\startlongtable
\begin{deluxetable}{lccc}
\tablecaption{Stellar and Tidal Energy Fluxes for Terrestrial Exoplanets \label{tab:demo}}
\tablewidth{0pt}
\tablehead{
\colhead{Planet} & \colhead{$F_{\rm star}$} & \colhead{$F_{\rm tide}$} & \colhead{$F_{\rm total}$} \\
\colhead{} & \colhead{(W\,m$^{-2}$)} & \colhead{(W\,m$^{-2}$)} & \colhead{(W\,m$^{-2}$)}
}
\startdata
55 Cnc e & $8.15\times10^{5}$ & $3.00\times10^{6}$ & $3.82\times10^{6}$ \\
K2-211 b & $7.00\times10^{5}$ & $2.51\times10^{7}$ & $2.58\times10^{7}$ \\
Kepler-323 b & $6.45\times10^{5}$ & $7.05\times10^{4}$ & $7.15\times10^{5}$ \\
KOI-94 b & $3.91\times10^{5}$ & $1.68\times10^{4}$ & $4.08\times10^{5}$ \\
TOI-500 b & $3.86\times10^{5}$ & $4.92\times10^{6}$ & $5.31\times10^{6}$ \\
Kepler-107 b & $3.64\times10^{5}$ & $4408$ & $3.68\times10^{5}$ \\
EPIC 249893012 b & $3.51\times10^{5}$ & $1670$ & $3.53\times10^{5}$ \\
Kepler-323 c & $2.37\times10^{5}$ & $2437$ & $2.40\times10^{5}$ \\
Kepler-107 c & $2.04\times10^{5}$ & $364$ & $2.05\times10^{5}$ \\
K2-265 b & $1.98\times10^{5}$ & $6.76\times10^{4}$ & $2.66\times10^{5}$ \\
GJ 367 b & $1.96\times10^{5}$ & $1.38\times10^{7}$ & $1.40\times10^{7}$ \\
TOI-1203 b & $1.94\times10^{5}$ & $495$ & $1.94\times10^{5}$ \\
Kepler-450 d & $1.93\times10^{5}$ & $18.97$ & $1.93\times10^{5}$ \\
K2-36 b & $1.76\times10^{5}$ & $1.72\times10^{5}$ & $3.47\times10^{5}$ \\
K2-111 b & $1.66\times10^{5}$ & $543$ & $1.67\times10^{5}$ \\
K2-138 b & $1.63\times10^{5}$ & $4353$ & $1.68\times10^{5}$ \\
Kepler-1876 b & $1.61\times10^{5}$ & $14.06$ & $1.61\times10^{5}$ \\
Kepler-23 b & $1.54\times10^{5}$ & $2.78$ & $1.54\times10^{5}$ \\
K2-38 b & $1.49\times10^{5}$ & $5361$ & $1.54\times10^{5}$ \\
TOI-1238 b & $1.46\times10^{5}$ & $1.52\times10^{7}$ & $1.54\times10^{7}$ \\
K2-19 d & $1.40\times10^{5}$ & $8.77\times10^{4}$ & $2.28\times10^{5}$ \\
HD 108236 b & $1.18\times10^{5}$ & $441$ & $1.18\times10^{5}$ \\
K2-226 b & $1.13\times10^{5}$ & $1.86\times10^{4}$ & $1.32\times10^{5}$ \\
Kepler-20 b & $1.12\times10^{5}$ & $2201$ & $1.14\times10^{5}$ \\
Kepler-1972 b & $1.10\times10^{5}$ & $3.75$ & $1.10\times10^{5}$ \\
Kepler-107 d & $1.07\times10^{5}$ & $9.53$ & $1.07\times10^{5}$ \\
HD 63433 d & $1.01\times10^{5}$ & $1027$ & $1.02\times10^{5}$ \\
K2-208 b & $9.00\times10^{4}$ & $1.96\times10^{4}$ & $1.10\times10^{5}$ \\
K2-147 b & $7.68\times10^{4}$ & $2.21\times10^{7}$ & $2.22\times10^{7}$ \\
Kepler-105 c & $7.63\times10^{4}$ & $2.57$ & $7.63\times10^{4}$ \\
TOI-733 b & $7.31\times10^{4}$ & $106$ & $7.32\times10^{4}$ \\
Kepler-1514 c & $7.29\times10^{4}$ & $51.23$ & $7.30\times10^{4}$ \\
K2-32 e & $7.26\times10^{4}$ & $83.83$ & $7.27\times10^{4}$ \\
Kepler-68 c & $6.94\times10^{4}$ & $4.45$ & $6.94\times10^{4}$ \\
Kepler-444 b & $6.41\times10^{4}$ & $112$ & $6.42\times10^{4}$ \\
Kepler-1972 c & $6.38\times10^{4}$ & $0.2562$ & $6.38\times10^{4}$ \\
K2-130 b & $5.95\times10^{4}$ & $2.25\times10^{4}$ & $8.20\times10^{4}$ \\
TOI-5788 b & $5.92\times10^{4}$ & $51.23$ & $5.92\times10^{4}$ \\
Kepler-20 e & $5.76\times10^{4}$ & $22.06$ & $5.76\times10^{4}$ \\
HD 15337 b & $5.72\times10^{4}$ & $303$ & $5.75\times10^{4}$ \\
HD 224018 b & $4.99\times10^{4}$ & $0.7793$ & $4.99\times10^{4}$ \\
Kepler-444 c & $4.70\times10^{4}$ & $246$ & $4.72\times10^{4}$ \\
K2-35 b & $4.58\times10^{4}$ & $2.01\times10^{4}$ & $6.60\times10^{4}$ \\
HD 23472 d & $4.37\times10^{4}$ & $82.25$ & $4.37\times10^{4}$ \\
TOI-512 b & $4.33\times10^{4}$ & $3.01$ & $4.33\times10^{4}$ \\
K2-199 b & $4.31\times10^{4}$ & $231$ & $4.34\times10^{4}$ \\
Kepler-11 b & $4.30\times10^{4}$ & $4.17$ & $4.30\times10^{4}$ \\
Kepler-128 c & $4.12\times10^{4}$ & $0.0309$ & $4.12\times10^{4}$ \\
K2-90 c & $3.84\times10^{4}$ & $1.67\times10^{4}$ & $5.51\times10^{4}$ \\
Kepler-85 b & $3.51\times10^{4}$ & $2.26$ & $3.51\times10^{4}$ \\
TOI-178 c & $3.37\times10^{4}$ & $0.0613$ & $3.37\times10^{4}$ \\
TOI-836 b & $3.26\times10^{4}$ & $653$ & $3.32\times10^{4}$ \\
LHS 1678 b & $3.21\times10^{4}$ & $2.98\times10^{4}$ & $6.20\times10^{4}$ \\
Kepler-444 d & $3.11\times10^{4}$ & $21.40$ & $3.11\times10^{4}$ \\
Kepler-102 b & $3.05\times10^{4}$ & $9.31$ & $3.05\times10^{4}$ \\
TOI-1749 b & $2.75\times10^{4}$ & $2.72\times10^{4}$ & $5.47\times10^{4}$ \\
K2-128 b & $2.35\times10^{4}$ & $962$ & $2.45\times10^{4}$ \\
Kepler-444 e & $2.31\times10^{4}$ & $2.37$ & $2.31\times10^{4}$ \\
LHS 1903 b & $2.27\times10^{4}$ & $505$ & $2.32\times10^{4}$ \\
K2-116 b & $2.12\times10^{4}$ & $22.48$ & $2.13\times10^{4}$ \\
HD 219134 c & $2.11\times10^{4}$ & $36.81$ & $2.12\times10^{4}$ \\
HIP 113103 b & $2.09\times10^{4}$ & $278$ & $2.11\times10^{4}$ \\
Kepler-102 c & $2.07\times10^{4}$ & $3.60$ & $2.07\times10^{4}$ \\
TOI-1235 b & $2.05\times10^{4}$ & $945$ & $2.15\times10^{4}$ \\
Kepler-85 c & $2.03\times10^{4}$ & $0.7282$ & $2.03\times10^{4}$ \\
HD 23472 e & $1.74\times10^{4}$ & $3.42$ & $1.74\times10^{4}$ \\
Kepler-444 f & $1.70\times10^{4}$ & $15.84$ & $1.70\times10^{4}$ \\
K2-203 b & $1.69\times10^{4}$ & $49.32$ & $1.69\times10^{4}$ \\
K2-81 b & $1.69\times10^{4}$ & $133$ & $1.70\times10^{4}$ \\
Kepler-37 b & $1.51\times10^{4}$ & $0.0265$ & $1.51\times10^{4}$ \\
K2-35 c & $1.46\times10^{4}$ & $728$ & $1.53\times10^{4}$ \\
HD 260655 b & $1.44\times10^{4}$ & $704$ & $1.51\times10^{4}$ \\
Kepler-1650 b & $1.42\times10^{4}$ & $406$ & $1.46\times10^{4}$ \\
TOI-1695 b & $1.34\times10^{4}$ & $8446$ & $2.18\times10^{4}$ \\
GJ 486 b & $1.33\times10^{4}$ & $9.21$ & $1.33\times10^{4}$ \\
Kepler-102 d & $1.25\times10^{4}$ & $4.41$ & $1.25\times10^{4}$ \\
TOI-1468 b & $1.24\times10^{4}$ & $453$ & $1.28\times10^{4}$ \\
Ross 176 b & $1.23\times10^{4}$ & $6004$ & $1.83\times10^{4}$ \\
Kepler-20 f & $1.22\times10^{4}$ & $0.1049$ & $1.22\times10^{4}$ \\
K2-136 b & $1.14\times10^{4}$ & $25.11$ & $1.14\times10^{4}$ \\
K2-266 c & $1.11\times10^{4}$ & $0.8451$ & $1.11\times10^{4}$ \\
K2-102 b & $1.09\times10^{4}$ & $9.06$ & $1.09\times10^{4}$ \\
Kepler-28 c & $9926$ & $1.25$ & $9927$ \\
HD 23472 f & $9825$ & $1.07$ & $9826$ \\
L 98-59 b & $8348$ & $387$ & $8735$ \\
Kepler-37 c & $8101$ & $0.0381$ & $8101$ \\
Kepler-1928 b & $7324$ & $2.00$ & $7326$ \\
K2-21 b & $7026$ & $35.26$ & $7062$ \\
GJ 1132 b & $6585$ & $811$ & $7397$ \\
TOI-270 b & $6519$ & $54.02$ & $6573$ \\
TOI-406 c & $6334$ & $236$ & $6571$ \\
HD 23472 b & $5973$ & $0.9505$ & $5974$ \\
HD 260655 c & $5479$ & $34.03$ & $5513$ \\
TOI-912 b & $5203$ & $4.21\times10^{4}$ & $4.73\times10^{4}$ \\
Kepler-296 c & $5070$ & $3899$ & $8969$ \\
TOI-771 b & $4924$ & $4325$ & $9249$ \\
LHS 1678 c & $4615$ & $74.54$ & $4689$ \\
L 98-59 c & $4348$ & $0.5579$ & $4349$ \\
HD 219134 f & $4207$ & $0.3223$ & $4207$ \\
TOI-776 b & $4125$ & $15.98$ & $4141$ \\
LTT 1445 A c & $3768$ & $9916$ & $1.37\times10^{4}$ \\
Kepler-102 f & $3389$ & $0.0161$ & $3389$ \\
Kepler-138 b & $3361$ & $0.0354$ & $3361$ \\
TOI-2096 b & $3223$ & $6984$ & $1.02\times10^{4}$ \\
LHS 1678 d & $3084$ & $15.88$ & $3100$ \\
HD 23472 c & $2977$ & $0.0437$ & $2977$ \\
Kepler-1651 b & $2771$ & $44.26$ & $2815$ \\
K2-72 b & $2666$ & $109$ & $2774$ \\
K2-136 d & $2407$ & $0.0698$ & $2407$ \\
Kepler-138 c & $2286$ & $0.0793$ & $2286$ \\
Kepler-296 b & $2205$ & $89.62$ & $2295$ \\
Kepler-409 b & $2082$ & $0.0208$ & $2082$ \\
LP 791-18 d & $1972$ & $0.3317$ & $1973$ \\
LHS 1140 c & $1774$ & $266$ & $2040$ \\
K2-129 b & $1744$ & $19.27$ & $1763$ \\
K2-72 d & $1706$ & $16.65$ & $1723$ \\
L 98-59 d & $1701$ & $0.2730$ & $1701$ \\
TOI-700 b & $1700$ & $1.71$ & $1702$ \\
HD 219134 d & $1603$ & $0.0140$ & $1603$ \\
TOI-6716 b & $1498$ & $1.74\times10^{4}$ & $1.89\times10^{4}$ \\
TOI-270 d & $1281$ & $0.0721$ & $1282$ \\
TOI-2096 c & $1259$ & $334$ & $1593$ \\
Kepler-138 d & $1149$ & $2.08\times10^{-3}$ & $1149$ \\
K2-3 c & $1085$ & $0.0397$ & $1085$ \\
TOI-5713 b & $824$ & $102$ & $926$ \\
TOI-1266 c & $811$ & $0.6589$ & $812$ \\
K2-72 c & $701$ & $0.8982$ & $702$ \\
LHS 1903 e & $698$ & $1.86\times10^{-3}$ & $698$ \\
Kepler-69 c & $664$ & $1.00\times10^{-5}$ & $664$ \\
TOI-6002 b & $605$ & $25.89$ & $631$ \\
Gliese 12 b & $573$ & $5.41$ & $578$ \\
Kepler-438 b & $543$ & $1.19\times10^{-3}$ & $543$ \\
K2-3 d & $492$ & $5.78\times10^{-3}$ & $492$ \\
Kepler-296 e & $482$ & $0.2565$ & $482$ \\
Kepler-440 b & $459$ & $0.0474$ & $459$ \\
TOI-700 e & $434$ & $7.16\times10^{-3}$ & $434$ \\
K2-72 e & $380$ & $0.1238$ & $380$ \\
TOI-700 d & $292$ & $1.18\times10^{-3}$ & $292$ \\
Kepler-442 b & $238$ & $5.53\times10^{-6}$ & $238$ \\
Kepler-296 f & $212$ & $0.0191$ & $212$ \\
LHS 1140 b & $144$ & $0.0408$ & $145$ \\
Kepler-186 f & $100$ & $1.84\times10^{-6}$ & $100$ \\
Kepler-441 b & $73.87$ & $1.89\times10^{-6}$ & $73.87$ \\
\enddata
\end{deluxetable}


\section{Case Studies}
\label{sec:cases}

The demographic analysis of Section~\ref{sec:demo} establishes the
prevalence of tidal heating across the terrestrial exoplanet
population and identifies the energy thresholds that govern planetary
climate transitions. Here, we apply this framework to case studies
spanning the three classification categories defined in
Section~\ref{sec:taxonomy}. We examine tidally dominated planets
(TOI-6716~b and TOI-912~b), historically compromised HZ planets
(TOI-700~d and LHS~1140~b), and a tidally compounded Venus analog
(GJ~12~b). For each, we compute the tidal heating rates and
circularization timescales using the formalism of
Section~\ref{sec:energy}. For the historically compromised and tidally
compounded cases, we additionally reconstruct the pre-MS stellar flux
histories (Figure~\ref{fig:prems}) and assess the implications for
atmospheric survival.


\subsection{Tidally Dominated Planets: TOI-6716 b and TOI-912 b}
\label{sec:tidal_dom}

Among the 143 terrestrial exoplanets in our sample, 17 have tidal
fluxes that individually exceed the runaway greenhouse threshold of
295~W/m$^2$. For these planets, tidal dissipation is the dominant
energy source in the planetary energy budget. We highlight two
examples that illustrate this regime.

TOI-6716~b is an Earth-sized planet ($R_p = 0.98 \pm
0.07$~$R_\oplus$, $M_p \approx 0.91$~$M_\oplus$) in a 4.72-day orbit
around a fully convective M~dwarf at a distance of 0.032~AU
\citep{scott2026}. At the best-fit eccentricity of $e = 0.88$, the
tidal flux of $\sim$17,400~W/m$^2$ exceeds the stellar flux of
$\sim$1,498~W/m$^2$ by more than an order of magnitude
(Table~\ref{tab:demo}), making tidal dissipation the overwhelmingly
dominant energy source. However, we note that the eccentricity is
reported as an upper limit ($e = 0.88^{+0.0}_{-0.88}$;
\citealt{scott2026}), meaning the data are consistent with a circular
orbit. If confirmed, the extreme eccentricity would make TOI-6716~b
one of the most tidally heated terrestrial planets known.

TOI-912~b is a sub-Neptune-mass planet ($R_p = 1.93 \pm
0.13$~$R_\oplus$, $M_p = 5.1 \pm 0.5$~$M_\oplus$) in an 8.65-day
orbit around an M2.5 dwarf \citep{lacedelli2026}. Its tidal flux of
$\sim$42,100~W/m$^2$ exceeds its stellar flux of
$\sim$5,200~W/m$^2$ by a factor of $\sim$8
(Table~\ref{tab:demo}). \citet{lacedelli2026} note that the eccentric
model is preferred by the Bayesian evidence, though additional RV data
would strengthen the eccentricity constraint. We also note that at
$R_p = 1.93$~$R_\oplus$, TOI-912~b lies near the empirical boundary
between rocky and volatile-rich compositions
\citep{rogers2015a,fulton2017}; if the planet harbors a significant
volatile envelope, the constant-$Q$ terrestrial tidal model adopted
here may not apply.

We caution that the equilibrium-tide formula (Equation~\ref{eq:tides})
is a truncated expansion in eccentricity and is most accurate for $e
\lesssim 0.3$. At $e = 0.88$ (TOI-6716~b), higher-order terms in the
tidal dissipation function become significant, and the quoted tidal
flux should be interpreted as an order-of-magnitude upper-bound
diagnostic rather than a precise prediction from a full high-$e$ tidal
evolution model.

These two systems exemplify the broader population of tidally
dominated planets in Figure~\ref{fig:exo}: short-period terrestrial
worlds whose orbital eccentricities, if confirmed, produce tidal
energy budgets that dwarf their stellar irradiation. The eccentricity
uncertainties underscore the critical need for dedicated RV follow-up
of these systems to confirm or refine the orbital parameters that
determine their tidal heating rates. Such follow-up would establish
whether the ``tidally dominated'' category is well-populated or
largely an artifact of poorly constrained eccentricities in the
current catalog.


\subsection{Historically Compromised HZ Planets: TOI-700 d and LHS 1140 b}
\label{sec:hist_comp}

The six sub-threshold planets in our sample (TOI-700~d, Kepler-442~b,
Kepler-296~f, LHS~1140~b, Kepler-186~f, and Kepler-441~b) all have
negligible present-day tidal heating, and the $a^{-15/2}$ distance
scaling of tidal dissipation means that no physical eccentricity can
bridge the gap to the runaway greenhouse threshold at their orbital
distances. However, as planets orbiting M~dwarfs, all six were
subjected to elevated stellar irradiation during the host star's
pre-MS contraction phase. We examine two examples whose host stars are
bright enough for both JWST atmospheric characterization and RV
orbital refinement, in contrast to the fainter Kepler targets in this
group.

\begin{figure*}
  \includegraphics[width=\linewidth]{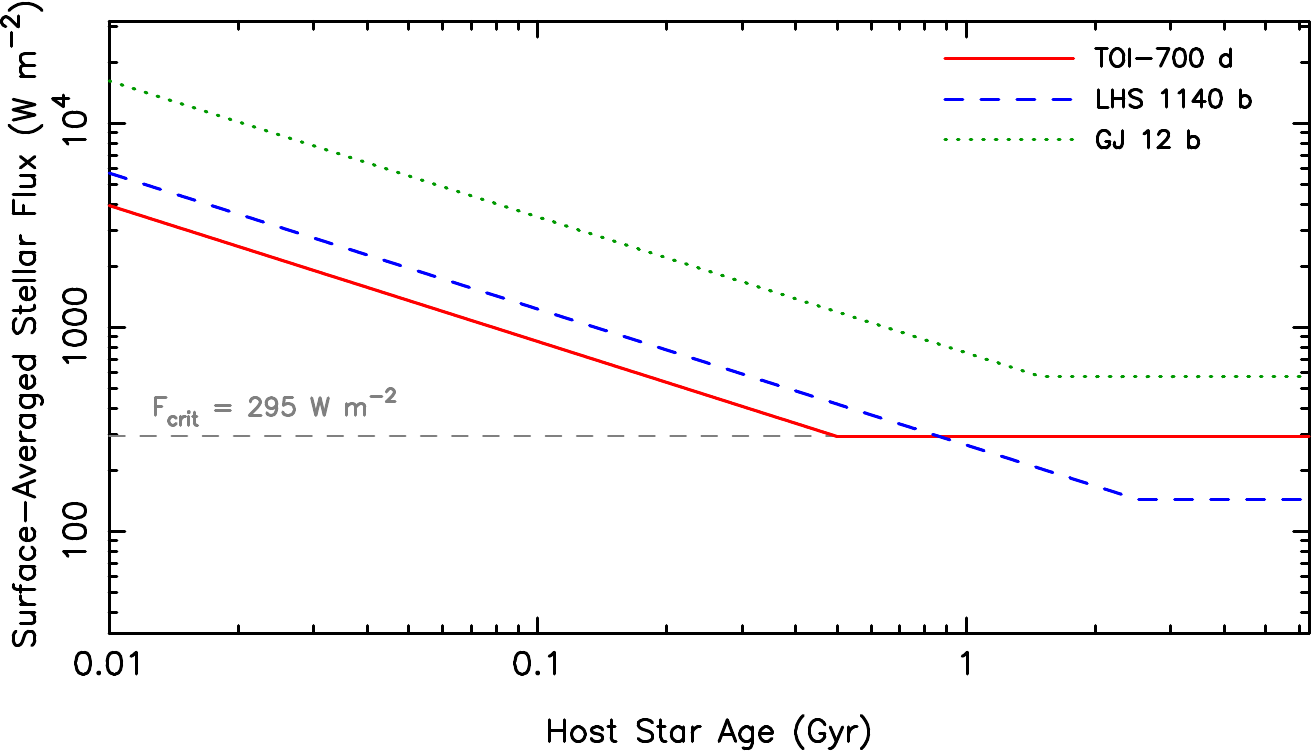}
  \caption{Surface-averaged stellar flux at the orbital distance of
    the historically compromised and tidally compounded case study
    planets as a function of host star age. The tidally dominated
    cases (TOI-6716~b and TOI-912~b) are omitted as their stellar
    fluxes exceed the runaway greenhouse threshold at all epochs
    regardless of pre-MS evolution. The
    luminosity evolution is approximated using Kelvin-Helmholtz
    contraction during the pre-MS phase, with contraction timescales
    of $\sim$0.5~Gyr (TOI-700, M2V), $\sim$1.5~Gyr (GJ~12, M3V),
    and $\sim$2.5~Gyr (LHS~1140, M4.5V), consistent with the
    evolutionary models of \citet{baraffe2015}. The horizontal dashed
    line marks the runaway greenhouse threshold at $F_{\rm crit} =
    295$~W/m$^2$. All three planets were irradiated well above the
    threshold during the host star's pre-MS phase. TOI-700~d drops
    below the threshold at $\sim$0.5~Gyr and plateaus at its
    present-day flux of $\sim$292~W/m$^2$ (visually coincident with
    but slightly below the threshold), LHS~1140~b crosses at
    $\sim$0.9~Gyr, while GJ~12~b remains above the threshold at all
    epochs.}
  \label{fig:prems}
\end{figure*}


\subsubsection{TOI-700 d}

TOI-700~d is an Earth-sized planet ($R_p = 1.16 \pm 0.06$~$R_\oplus$,
$M_p = 2.40^{+0.49}_{-0.52}$~$M_\oplus$) in a 37.42-day orbit around
an M2V dwarf ($M_\star = 0.416$~$M_\odot$, $R_\star =
0.42$~$R_\odot$) at a distance of 0.161~AU
\citep{gilbert2020b,gilbert2023}. The star is notably quiescent, with
no white-light flares detected across 11 TESS sectors. TOI-700~d
resides within the conservative HZ and receives a stellar flux of
$\sim$0.86~$S_\oplus$.

From Table~\ref{tab:demo}, TOI-700~d has a surface-averaged stellar
flux of $F_{\rm star} \approx 292$~W/m$^2$, placing it $\sim$3~W/m$^2$
below the runaway greenhouse threshold of 295~W/m$^2$. It is the most
marginal HZ planet in our sample. Its measured eccentricity of $e =
0.047^{+0.054}_{-0.030}$ \citep{gilbert2020b} yields a tidal
coefficient of $C_{\rm tidal} \approx 0.93$~W/m$^2$ per $e^2$,
resulting in a present-day tidal flux of $F_{\rm tide} \approx
0.002$~W/m$^2$, negligible compared to the 3~W/m$^2$ margin. The
relatively long orbital period (37.42~days) and large semi-major axis
place TOI-700~d in a regime where tidal effects scale as $a^{-15/2}$
to extremely small values, meaning that tidal heating alone cannot
bridge the gap to the runaway greenhouse threshold at any physical
eccentricity ($e_{\rm crit} \approx 1.8$). The circularization
timescale is $\tau_{\rm circ} \approx 7000$~Gyr, vastly exceeding any
plausible system age, so the eccentricity is effectively permanent.

The significance of TOI-700~d for the tidal Venus framework lies not
in the present-day tidal contribution but in the pre-MS history of
the system. TOI-700 is an M2 dwarf whose pre-MS contraction phase
lasts $\sim$0.5--1~Gyr \citep{baraffe2015}. During this epoch, the
stellar luminosity was a factor of $\sim$3--10 above the present-day
value \citep{ramirez2014c}, meaning that the stellar flux at
TOI-700~d's orbit would have been $\sim$876--2920~W/m$^2$, far above
the runaway greenhouse threshold, regardless of albedo
(Figure~\ref{fig:prems}). For
$\sim$0.5--1~Gyr, TOI-700~d was unambiguously in a runaway greenhouse
regime from stellar irradiation alone. The fact that it now sits
$\sim$3~W/m$^2$ below the threshold means that its present-day
habitability depends entirely on whether the atmosphere and water
inventory survived the pre-MS epoch. The four-planet architecture of
the TOI-700 system \citep{gilbert2023} maintains eccentricities of
$0.03 < e < 0.1$ through mutual secular perturbations, providing a
modest but persistent tidal contribution throughout the system
history. TOI-700~d thus represents a Category~3 (historically
compromised) planet in the taxonomy of Section~\ref{sec:taxonomy}: a
world whose present-day position near the runaway greenhouse boundary
belies a history of extreme irradiation.


\subsubsection{LHS 1140 b}
\label{sec:lhs1140}

LHS~1140~b is a super-Earth ($R_p = 1.730 \pm 0.025$~$R_\oplus$, $M_p
= 5.60 \pm 0.19$~$M_\oplus$) in a 24.74-day orbit around an M4.5V
dwarf ($M_\star = 0.1844$~$M_\odot$, $R_\star = 0.2159$~$R_\odot$) at
a distance of 0.0946~AU \citep{dittmann2017a,cadieux2024a}. The system
is old ($\gtrsim 5$~Gyr; \citealt{dittmann2017a,cadieux2024a}), and
the host star is inactive, with no observed flares. LHS~1140~b resides
deep within the conservative HZ and receives a stellar flux of
$\sim$0.43~$S_\oplus$. From Table~\ref{tab:demo}, LHS~1140~b has a
surface-averaged stellar flux of $F_{\rm star} \approx 144$~W/m$^2$,
well below the runaway greenhouse threshold. Its eccentricity is
consistent with zero ($e < 0.043$ at 95\% confidence;
\citealt{cadieux2024a}), and the tidal coefficient of $C_{\rm tidal}
\approx 22$~W/m$^2$ per $e^2$ yields a negligible present-day tidal
flux. An eccentricity of $e \approx 0.30$ would be required to reach
the extreme volcanism threshold of 2~W/m$^2$, and the circularization
timescale of $\tau_{\rm circ} \approx 228$~Gyr means any historical
eccentricity would persist to the present day.

Like TOI-700~d, LHS~1140~b's relevance to the tidal Venus framework is
primarily historical. LHS~1140 is an M4.5 dwarf whose pre-MS
contraction phase lasts $\sim$1--3~Gyr \citep{baraffe2015}. During
this epoch, the stellar flux at LHS~1140~b's orbit would have been a
factor of $\sim$3--10 higher, or $\sim$432--1440~W/m$^2$, well above
the runaway greenhouse threshold (Figure~\ref{fig:prems}). LHS~1140~b
was therefore exposed to above-threshold irradiation for an extended
period during the early history of the system.

However, LHS~1140~b differs from TOI-700~d in one critical respect:
its mass. At 5.6~$M_\oplus$, LHS~1140~b has a surface gravity of
$\sim$18~m/s$^2$ and an escape velocity of $\sim$15~km/s,
substantially higher than Earth. This potentially results in the
planet having superior atmospheric retention against both thermal escape
and stellar-wind stripping \citep{zahnle2017}. Recent analyses by
\citet{cadieux2024a} indicate that LHS~1140~b's density of
$\sim$5.9~g/cm$^3$ is consistent with either a rocky interior with a
thin H$_2$/He atmosphere or a water world with a significant ice/water
mass fraction. JWST/NIRISS transmission spectroscopy has ruled out
H$_2$-rich atmospheres at $>$10$\sigma$ and revealed tentative
evidence (2.3$\sigma$) of Rayleigh scattering consistent with an
N$_2$-dominated atmosphere \citep{cadieux2024a}, suggesting that
LHS~1140~b is either airless or surrounded by a high mean molecular
weight atmosphere.

LHS~1140~b therefore represents the most stringent observational test
of whether massive terrestrial planets can survive the M~dwarf pre-MS
gauntlet with their atmospheres intact. If JWST confirms an
atmosphere, it constrains the effectiveness of pre-MS irradiation at
stripping volatiles from high-mass terrestrial worlds, providing a
direct calibration point for the Category~3 (historically compromised)
classification. If no atmosphere is detected, it would suggest that
even 5.6~$M_\oplus$ is insufficient to retain volatiles through a
multi-Gyr above-threshold epoch, consistent with predictions of
extreme atmospheric loss around M~dwarfs
\citep{luger2015b,tian2015d,vanlooveren2024} and with profound
implications for the habitability of all M~dwarf terrestrial planets.


\subsection{GJ 12 b: Tidal Compounding in a Venus Analog}
\label{sec:gj12b}

GJ~12~b represents the most data-rich case study in our sample and
exemplifies a Category~1 (flux-driven Venus analog) planet with
significant tidal compounding. Unlike TOI-700~d and LHS~1140~b, whose
stellar fluxes fall below the runaway greenhouse threshold, GJ~12~b's
stellar irradiation of $\sim$1.68~$S_\oplus$ already exceeds the
threshold for albedos below $\sim$0.49. Tidal heating does not change
the classification but compounds the Venus-like state through Super-Io
volcanism and an independent geochemical pathway to a runaway
greenhouse.


\subsubsection{System Properties and Habitable Zone}
\label{sec:system}
 
GJ~12 is a relatively close M3.0 dwarf star with a single known
planet, the properties of which are shown in Table~\ref{tab:system}.
GJ~12 is classified as among the least active M dwarfs known, with an
X-ray-to-bolometric luminosity ratio $\log(L_X/L_{\rm bol}) \lesssim
-5.5$ \citep{kuzuhara2024}. Kinematic age estimates derived from the
UVW space velocities of GJ~12 return an age of $\gtrsim$2.8~Gyr,
consistent with an older, most likely super-solar age and thin-disk
membership \citep{dholakia2024}. This means that GJ~12 has long since
completed its pre-main-sequence (pre-MS) contraction phase, implying
that GJ~12~b has been exposed to several Gyrs of evolving stellar
irradiation and, potentially, tidal heating.

The planet was discovered by \citet{kuzuhara2024} and
\citet{dholakia2024} using TESS photometry in combination with
ground-based follow-up transit observations and RV
measurements. \citet{brady2025b} and \citet{turner2026a} provided mass
measurements from follow-up RV observations, the results of which
allowed the bulk density to be calculated, and shown to be consistent
with a predominantly rocky interior comparable to Earth and
Venus. These works also report a best-fit eccentricity of $e = 0.24
\pm 0.11$, noting that the data do not statistically rule out $e =
0$. We calculated a revised semi-major axis for the planet of $a =
0.0678$~AU using the \citet{turner2026a} stellar mass of
$0.255$~$M_\odot$ (see Table~\ref{tab:system}), both of which are
slightly higher than those provided by \citet{kuzuhara2024}.

\begin{deluxetable}{lcc}
\tablecaption{GJ~12 System Parameters \label{tab:system}}
\tablewidth{0pt}
\tablehead{
\colhead{Parameter} & \colhead{Value} & \colhead{Reference}
}
\startdata
\multicolumn{3}{c}{\textit{Stellar Parameters}} \\
\hline
Distance (pc) & $12.162 \pm 0.005$ & K24, D24 \\
$T_{\rm eff}$ (K) & $3328 \pm 78$ & T26 \\
$M_\star$ ($M_\odot$) & $0.255 \pm 0.013$ & T26 \\
$R_\star$ ($R_\odot$) & $0.265 \pm 0.012$ & T26 \\
$L_\star$ ($L_\odot$) & $7.73 \times 10^{-3}$ & This work \\
$\log(L_X/L_{\rm bol})$ & $\lesssim -5.5$ & K24 \\
OHZ (inner, AU) & 0.072 & This work \\
CHZ (inner, AU) & 0.091 & This work \\
CHZ (outer, AU) & 0.179 & This work \\
OHZ (outer, AU) & 0.189 & This work \\
\hline
\multicolumn{3}{c}{\textit{Planetary Parameters}} \\
\hline
$P$ (days) & $12.761421 \pm 0.000047$ & T26 \\
$a$ (AU) & $0.0678$ & This work \\
$R_p$ ($R_\oplus$) & $0.93 \pm 0.06$ & T26 \\
$M_p$ ($M_\oplus$) & $0.95^{+0.29}_{-0.30}$ & T26 \\
$\rho_p$ (g\,cm$^{-3}$) & $6.4 \pm 2.4$ & T26 \\
$T_{\rm eq}$ (K) & $317 \pm 8$ & T26 \\
$e$ & $0.24 \pm 0.11$ & B25, T26 \\
$\bar{S}/S_\oplus$ ($e=0$) & $\sim$1.68 & This work \\
TSM & $\sim$70 & K24 \\
\enddata
\tablecomments{References: K24 = \citet{kuzuhara2024}; D24 =
  \citet{dholakia2024}; B25 = \citet{brady2025b}; T26 =
  \citet{turner2026a}.}
\end{deluxetable}

\begin{figure}
  \includegraphics[width=\linewidth]{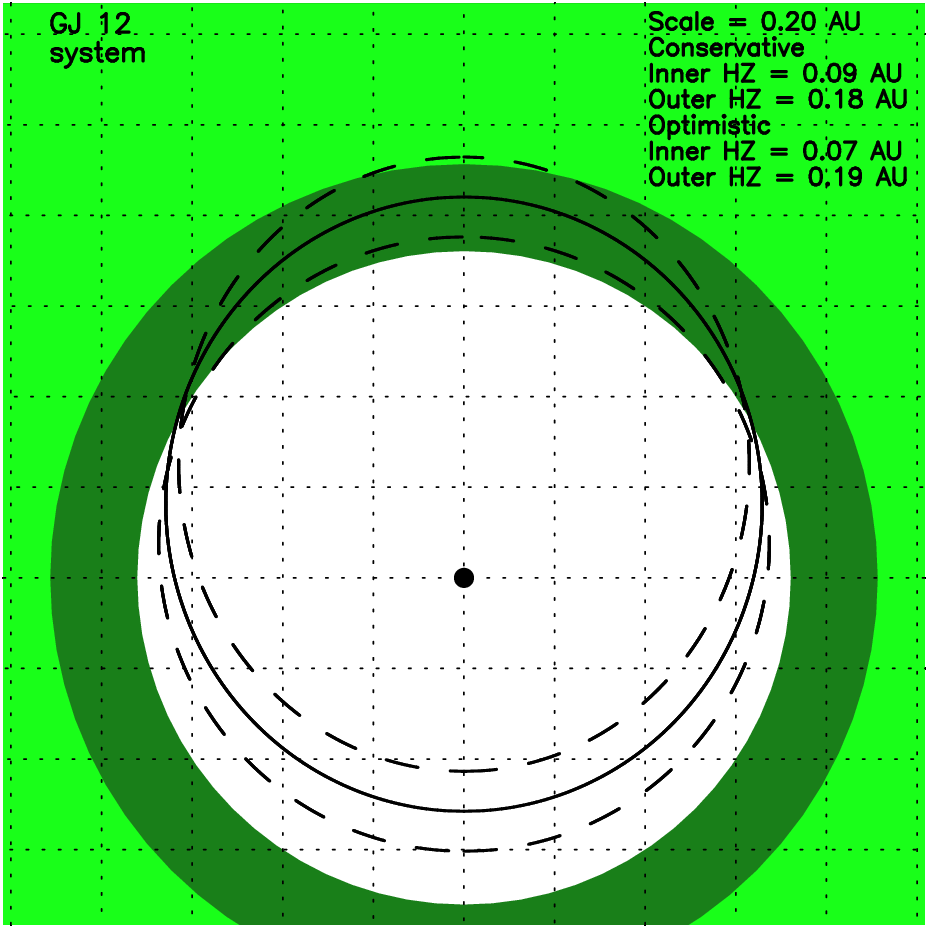}
  \caption{HZ and planetary orbits in the GJ~12 system. The extent of
    the HZ is shown in green, where light green and dark green
    indicate the CHZ and OHZ, respectively. The solid line shows the
    orbit with the nominal eccentricity of 0.24, and the dashed lines
    represent 1$\sigma$ boundaries \citep{turner2026a}. The scale of
    the figure is 0.2~AU along each side with a 0.02~AU grid.}
  \label{fig:hz}
\end{figure}

We calculated the HZ of the system, adopting the methodology described
by \citet{kopparapu2013a,kopparapu2014}. We divide the HZ into the
conservative HZ (CHZ) and optimistic HZ (OHZ), the latter of which is
based upon assumptions regarding the prevalence of surface liquid
water for Venus and Mars \citep{kane2016c}. For GJ~12, the HZ regions
span 0.091--0.179~AU and 0.072--0.189~AU for the CHZ and OHZ,
respectively. The inner components of the CHZ (light green) and OHZ
(dark green) are shown in Figure~\ref{fig:hz}. The solid line shows
the orbit of GJ~12~b using the orbital parameters from
Table~\ref{tab:system}, and the dashed lines indicate orbits
representative of the 1$\sigma$ uncertainties on the orbital
eccentricity. For each of these scenarios, the planet generally
maintains an apastron passage, where the orbital motion is the
slowest, within the OHZ.
 
\subsubsection{Climate Models for Temperate and Runaway Greenhouse Scenarios}
\label{sec:gcm}
 
To test the viability of several climate states of GJ~12~b, we
utilized the Generic Planetary Climate Model (Generic-PCM); a
3-dimensional model that computes coupled dynamical and physical
processes to predict the potential climate state of a planet. These
calculations are based on the system parameters as well as the
prescribed initial surface and atmospheric conditions. The model
implements a generalized radiative transfer routine
\citep{wordsworth2010b} and correlated-k tables from kspectrum
\citep{eymet2016}. Many different atmospheres and climates for planets
within our solar system
\citep[e.g.,][]{wordsworth2013a,lebonnois2016}, and exoplanets
\citep[e.g.,][]{leconte2013b,turbet2016,fauchez2019,quirino2023} have
been simulated using the Generic-PCM.

Our simulations adopt the assumption of a circular orbit (constant
stellar flux) to determine the baseline flux outcome without the
addition of tidal heating. Stellar and planetary parameters for the
simulations are adopted from Table~\ref{tab:system}. The stellar
spectrum data input is based on the BT-Settl stellar model grid
\citep{allard2014}, which was selected to match the properties of
GJ~12 (same effective temperature, surface gravity, and
metallicity). In both scenarios, GJ~12~b is assumed to be tidally
locked.

We adopted two separate climate scenarios for GJ~12~b to assess the
planet's climate stability. The first scenario simulated GJ~12~b under
temperate, Earth-similar conditions. The atmosphere was assumed to be
N$_2$-dominated with pre-industrial levels of CO$_2$ (N$_2$ = 99.97\%,
CO$_2$ = 280~ppm, H$_2$O acts as a variable gas), and a surface
pressure of $\sim$1~bar. The model resolution is $64 \times 48$
(longitude-latitude), with 30 atmospheric layers extending to the
top-of-atmosphere (TOA) pressure of 4~Pa. The second scenario
simulated GJ~12~b under Venus-similar conditions, adopting a
CO$_2$-dominated Venus-like atmosphere (CO$_2$ = 99.96\%, SO$_2$ =
180~ppm, H$_2$O = 31~ppm, CO = 12~ppm, OCS = 51~ppm). The model
included radiatively active, globally distributed Venus-like sulfuric
acid clouds, following \citet{haus2014}. The surface pressure was set
to $\sim$92 bar, with a model resolution of $64 \times 48$
(longitude-latitude), and with 50 atmospheric layers extending to a
TOA pressure of 5~Pa. Each model scenario was run until either the
model failed due to exceeding the correlated-k dataset temperature
range, or achieved radiative balance with a TOA convergence criterion
of $< 1$~W/m$^2$ \citep{turbet2023}. The simulations were adapted
using ``Earth slab-ocean'' simulation start files (with the slab ocean
parameter disabled) for the GJ~12~b temperate case
\citep{codron2012,charnay2013}, and the ``TRAPPIST-1 c with Venus-like
conditions simulation'' start files for the Venus-like case
\citep{quirino2023} that are also available from the Generic-PCM
website.

\begin{figure*}
  \centering
  \includegraphics[width=0.9\linewidth]{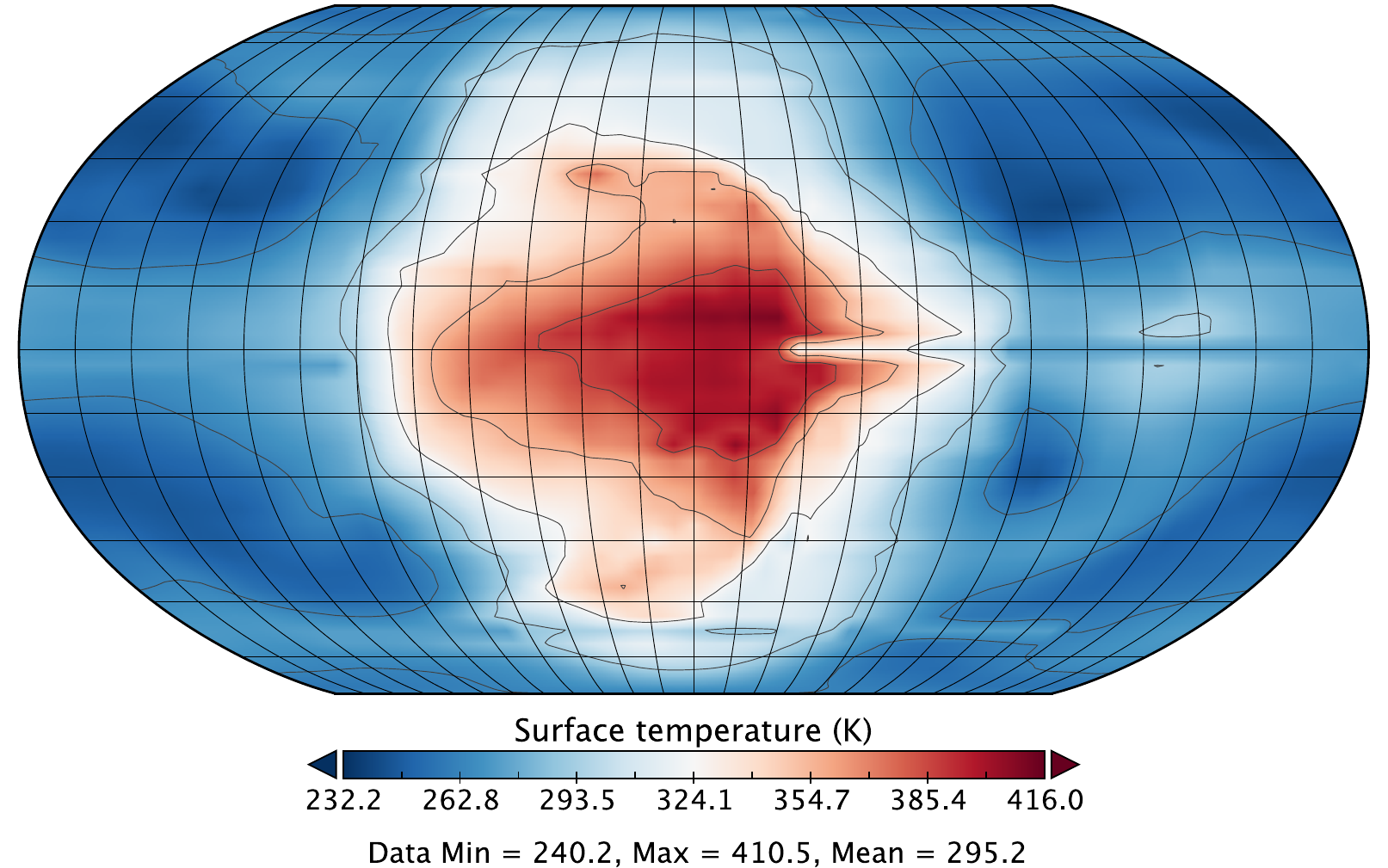} \\
  \includegraphics[width=0.9\linewidth]{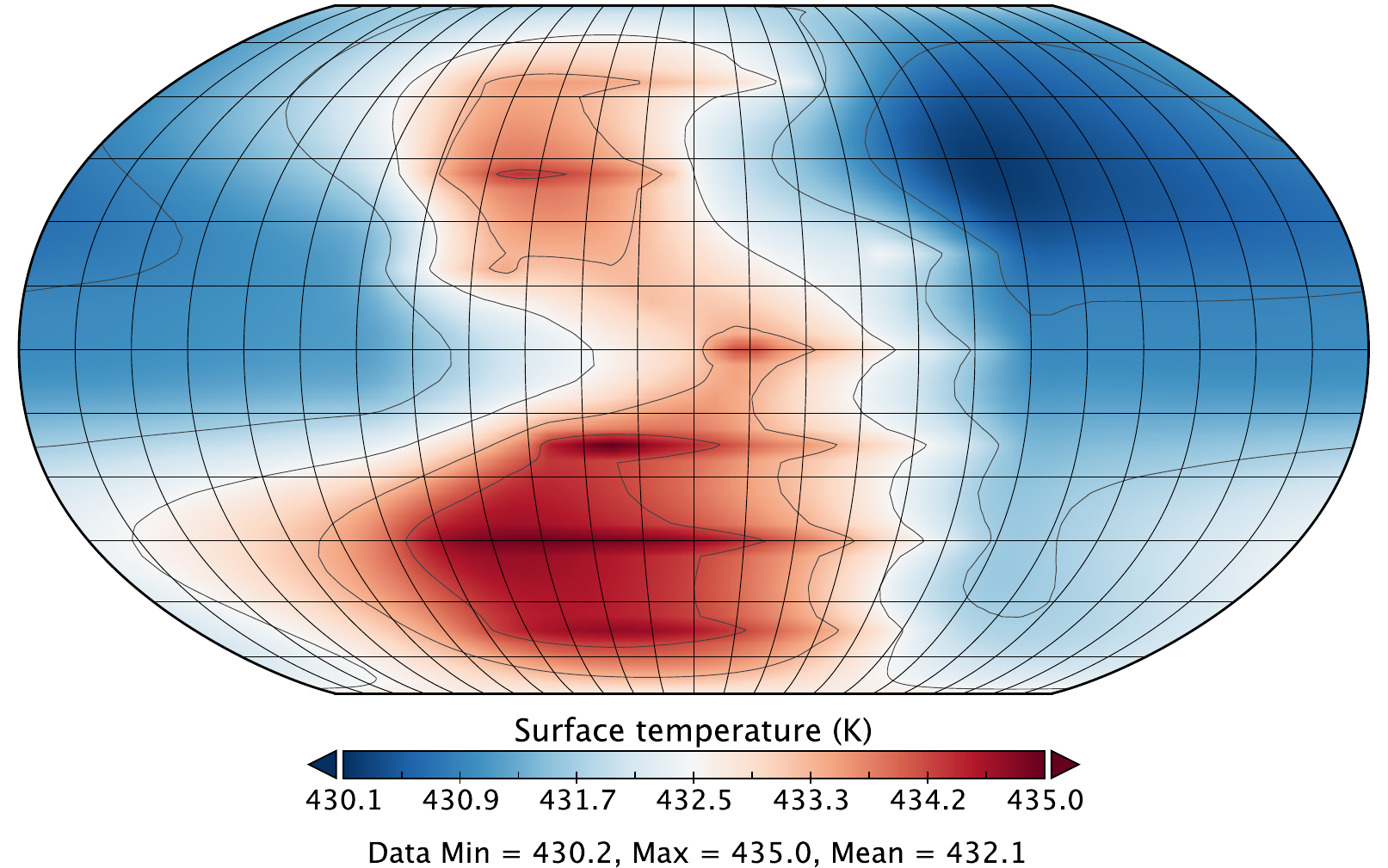}
  \caption{Surface temperature outputs from the Generic-PCM using the
    temperate climate scenario (top panel) and the Venus climate scenario
    (bottom panel). The temperate model was not able to achieve
    radiative balance due to escalating temperatures, whereas the Venus
    model achieved radiative balance. Note the different color scales
    between the two panels.}
  \label{fig:gcm}
\end{figure*}

The resulting surface temperature maps are shown in
Figure~\ref{fig:gcm} for the temperate scenario (top panel) and Venus
scenario (bottom panel). The temperate model was not able to achieve
radiative balance due to the rapidly rising temperatures exceeding the
correlated-k dataset. The surface temperature map shown in the top
panel of Figure~\ref{fig:gcm} thus represents the final state of the
model prior to failure, and exhibits a vast temperature range
consistent with high temperatures at the sub-stellar point. The Venus
model, however, was able to achieve radiative balance and converge to
a final state, as shown in the bottom panel of
Figure~\ref{fig:gcm}. The average surface temperature from that model
was 430~K, consistent with the lower incident flux received by GJ~12~b
compared to the 1.91~$S_\oplus$ of modern Venus. Together, these
models add credence to the notion that GJ~12~b may indeed be in a
stable post-runaway greenhouse state given the prescribed initial
conditions.

The 3D simulations presented here provide several advances beyond
existing 1D treatments of the runaway greenhouse
\citep{goldblatt2013,boer2025}. First, GJ~12~b is a tidally locked
planet orbiting an M3 dwarf, and the spectral energy distribution,
substellar geometry, and day--night heat redistribution fundamentally
alter the climate dynamics in ways that 1D models cannot capture.
Three-dimensional GCMs have shown that substellar cloud formation can
raise the runaway greenhouse threshold for slowly rotating planets
\citep{leconte2013c,yang2014b,boukrouche2025}, making 3D verification
essential for specific systems. Second, the surface temperature maps
(Figure~\ref{fig:gcm}) provide spatially resolved predictions that can
be tested against JWST thermal phase curves. Third, the quantitative
convergence of the Venus-like simulation to a mean surface temperature
of 430~K provides a system-specific prediction that does not exist in
the 1D literature for GJ~12~b.


\subsubsection{Energy Budget and the Albedo Threshold}
\label{sec:energy_budget}
 
We compute the total effective flux $F_{\rm total}$ for GJ~12~b as a
function of orbital eccentricity using Equations~(\ref{eq:flux}) and
(\ref{eq:tides}). The incident stellar flux at the semi-major axis is:
\begin{equation}
    \frac{\bar{S}(e=0)}{S_\oplus} = \frac{L_\star}{L_\odot}
    \left(\frac{1~\mathrm{AU}}{a}\right)^2 \approx 1.68,
    \label{eq:insolation_gj12b}
\end{equation}
\noindent placing GJ~12~b at a flux intermediate between Earth
(1.0\,$S_\oplus$) and Venus ($\sim$1.91\,$S_\oplus$). This insolation
already exceeds the conservative 1D runaway greenhouse threshold of
$S_{\rm RG} = 1.06\,S_\oplus$ \citep{kopparapu2013a}, meaning that
under a cloud-free, low-albedo atmosphere, GJ~12~b is unlikely to be
habitable on the basis of stellar irradiation alone. The planet is
therefore not necessarily a case where tidal heating induces a
Venus-like state from an otherwise temperate baseline. Rather, the
planet may have experienced an interplay between its albedo and tidal
heating, increasing the role of tidal heating during earlier, more
dynamically active epochs in the system's history.

The analysis below treats the Bond albedo parametrically, mapping the
threshold behavior as a function of albedo and eccentricity. This
approach is standard in the tidal heating and VZ literature
\citep{barnes2013a,kane2014e,heller2015b} and is complementary to the
self-consistent 3D climate simulations presented in
Section~\ref{sec:gcm}. The parametric treatment is physically
motivated because it reveals the conditions under which GJ~12~b
crosses the runaway greenhouse threshold, without requiring specific
knowledge of atmospheric composition, cloud microphysics, or volatile
inventory, all of which are unconstrained for this planet. The 3D
simulations, which include cloud and surface physics within the
Generic-PCM framework, provide an independent and self-consistent
verification for two specific atmospheric compositions.

The fate of GJ~12~b's atmosphere depends on its Bond albedo, $A$. The
absorbed stellar flux is:
\begin{equation}
    S_{\rm abs}(A) = (1 - A) \cdot \frac{\bar{S}}{S_\oplus}.
\end{equation}
For a Venus-like albedo ($A \sim 0.7$), $S_{\rm abs} \approx
0.5$~$S_\oplus$, which is well below any runaway greenhouse threshold.
However, we emphasize that a Bond albedo of $\sim$0.7 in the Solar
System is achieved specifically by Venus's thick sulfuric acid cloud
deck, which is itself the product of a dense CO$_2$ greenhouse
atmosphere with surface temperatures of $\sim$735~K
\citep{mahapatra2023}. A high albedo of this origin does not imply a
temperate surface; it implies a planet that is already in a Venus-like
state. The physically relevant question for habitability is whether
intermediate albedos ($A \approx 0.3$--0.5), arising from mechanisms
such as water clouds, ice coverage, or reflective aerosols, could
shield a temperate atmosphere from the runaway greenhouse. For an
Earth-like albedo ($A \approx 0.3$), $S_{\rm abs} \approx
1.18$~$S_\oplus$, which lies above the conservative threshold
\citep{kopparapu2013a} but below the 3D GCM threshold of
$\sim$1.4~$S_\oplus$ for slowly rotating planets
\citep{leconte2013c}. In principle, then, the question of whether
GJ~12~b is, or has ever been, temperate hinges on whether it has
developed or retained a sufficiently reflective atmosphere.

Unlike stellar irradiation, tidal energy is deposited directly in the
planetary interior and must be radiated as internal heat flux,
regardless of the planet's atmospheric albedo. It therefore
contributes to $F_{\rm total}$ in a way that cannot be mitigated by a
reflective cloud layer. For tidal dissipation, adopting the system
properties from Table~\ref{tab:system}, as well as $k_2 = 0.3$ and $Q
= 50$, we find:
\begin{equation}
    F_{\rm tide} = C_{\rm tidal} \cdot e^2,
\end{equation}
where the coefficient $C_{\rm tidal} \approx 93.9$~W/m$^2$ for
GJ~12~b, consistent with the value $F_{\rm tide} = 5.41$~W/m$^2$ at $e
= 0.24$ shown in Table~\ref{tab:demo}. The total effective surface
flux is therefore:
\begin{equation}
    F_{\rm total}(e) \approx 573 + 93.9 \, e^2
    \label{eq:stot_gj12b}
\end{equation}
The stellar flux of $\sim$573~W/m$^2$ already exceeds the runaway
greenhouse limit of 295~W/m$^2$ for albedos $A \lesssim 0.49$. The
threshold albedo above which stellar flux alone cannot trigger a
runaway greenhouse is $A_{\rm thresh} = 1 - F_{\rm crit}/F_{\rm star}
\approx 0.49$. For a planet whose temperate state requires an albedo
above this threshold, tidal heating provides the additional
push. Adopting $A = 0.5$, for which $S_{\rm abs} \approx 287$~W/m$^2$
(just below the runaway threshold) the critical eccentricity is:
\begin{equation}
    e_{\rm crit} = \sqrt{\frac{F_{\rm crit} - F_{\rm abs}}{C_{\rm
          tidal}}} = \sqrt{\frac{295 - 287}{93.9}} \approx 0.30.
    \label{eq:ecrit}
\end{equation}
An eccentricity of $e \approx 0.30$ lies within the measured
uncertainty range of $e = 0.24 \pm 0.11$
\citep{brady2025b,turner2026a}, meaning tidal heating at the measured
eccentricity can plausibly close the margin for atmospheres with
moderate high albedo. For a Venus-like albedo ($A \sim 0.7$, which as
noted above is itself indicative of a non-temperate state), the
absorbed stellar flux ($\sim$172~W/m$^2$) is substantially below the
runaway threshold, and the eccentricity required to bridge the gap
through tidal heating alone is the unphysical value of $e_{\rm crit} =
\sqrt{(295 - 172)/93.9} \approx 1.1$. Consequently, tidal heating
plays a decisive role for GJ~12~b primarily in the moderate-albedo
regime ($A \approx 0.5$), where it tips a marginally sub-threshold
planet into a runaway state at realistic eccentricities.

To assess the sensitivity of these results to the eccentricity
uncertainty, we propagate the measured $e = 0.24 \pm 0.11$ through the
tidal heating calculation. At the nominal eccentricity, $F_{\rm tide}
= 5.4$~W/m$^2$; at the 1$\sigma$ upper bound ($e = 0.35$), $F_{\rm
tide} = 11.5$~W/m$^2$; and at the 1$\sigma$ lower bound ($e = 0.13$),
$F_{\rm tide} = 1.6$~W/m$^2$. Even at the lower bound, the tidal flux
approaches the extreme volcanism threshold of 2~W/m$^2$. For $e = 0$,
the tidal contribution vanishes entirely, but the stellar flux of
$\sim$573~W/m$^2$ still exceeds the runaway greenhouse limit for any
albedo below $\sim$0.49. More critically, the pre-main-sequence
argument presented in Section~\ref{sec:prems} provides a pathway to a
Venus-like outcome that is independent of the present-day eccentricity:
regardless of whether the current eccentricity is 0.24 or zero, the
planet was exposed to extreme combined irradiation and tidal heating
during the first $\sim$1--3~Gyr of the system's history. The current
eccentricity constrains the present tidal contribution but does not
determine the planet's evolutionary fate.
 

\subsubsection{The Pre-Main-Sequence Tidal Greenhouse Epoch}
\label{sec:prems}
 
Thus far, our examination of GJ~12~b has addressed the present-day
energy budget. However, the old age of the system ($\gtrsim$2.8~Gyr;
\citealt{dholakia2024,turner2026a}) opens an additional pathway to a
Venus-like outcome, which is through the coupled effect of elevated
pre-MS stellar luminosity and tidal heating during the early history
of the system.
 
M dwarfs are significantly more luminous during their pre-MS
contraction phase which, for a star of GJ~12's mass, lasts
$\sim$1--3~Gyr before settling onto the main sequence
\citep{ramirez2014c}. During this epoch, the stellar luminosity can
exceed the present-day value by factors of several, and the HZ lies at
substantially larger orbital distances \citep{baraffe2015}. A planet
at GJ~12~b's current semi-major axis of 0.0678~AU would therefore have
received a considerably higher stellar flux than it does today,
potentially well in excess of any runaway greenhouse threshold,
regardless of albedo (Figure~\ref{fig:prems}). Moreover, the elevated EUV and X-ray environment
during the pre-MS active phase accelerates the photodissociation of
water vapor and hydrogen escape \citep{luger2015b}, with water loss
timescales of $\tau_{\rm loss} \sim 10^7$--$10^9$~yr for M dwarf EUV
fluxes and Earth-like water inventories.
 
If GJ~12~b additionally harbored a non-zero orbital eccentricity
during this early epoch, before several Gyr of tidal damping have
acted, the tidal contribution to the total energy budget would have
been correspondingly larger. The tidal circularization timescale of
$\tau_{\rm circ} \approx 9.7$~Gyr (Section~\ref{sec:dynamics}) is far
longer than the pre-MS phase itself, meaning that any eccentricity
present during the pre-MS epoch persists essentially unchanged to the
present day and would have driven sustained tidal heating throughout
that entire early period. During the first $\sim$1--3~Gyr of the
system's history, the combined energy input could have established a
runaway greenhouse and driven complete water loss. In this scenario,
the present-day state of GJ~12~b is not the product of current
conditions, but the fossil record of a tidal and irradiative
catastrophe that occurred Gyrs ago. The planet's current insolation of
$\sim$1.68\,$S_\oplus$ and its position relative to the present-day
runaway threshold may then be a secondary consideration.
 

\subsubsection{Outgassing Pathway to a Runaway Greenhouse}
\label{sec:volcanic}

The demographic analysis in Section~\ref{sec:demo} showed that
GJ~12~b's current tidal flux of 5.4~W/m$^2$ places it above the
extreme volcanism threshold of 2~W/m$^2$. This classifies GJ~12~b as a
candidate Super-Io world: a planet experiencing tidal dissipation
sufficient to drive geologically dominant volcanic activity at a rate
orders of magnitude above that of present-day Earth
\citep{heller2013a,driscoll2015}. Yet the tidal flux falls far short
of the runaway greenhouse flux of 295~W/m$^2$, meaning tidal heating
alone cannot drive a runaway greenhouse through direct thermal
forcing.

However, there is a geochemically mediated pathway that connects
extreme volcanism to a runaway greenhouse, independent of direct
thermal forcing. At Super-Io levels of volcanic activity, a planet
releases enormous quantities of CO$_2$, SO$_2$, and H$_2$O into the
atmosphere through mantle outgassing. On present-day Earth, volcanic
CO$_2$ outgassing is balanced by silicate weathering and carbonate
precipitation on timescales of $\sim$$10^5$--$10^6$ yr; this
carbon-silicate cycle stabilizes the climate against long-term warming
\citep{walker1981,berner1983b,schaefer2015,oosterloo2021}. However,
for a planet with GJ~12~b's stellar flux ($\sim$573~W/m$^2$
surface-average, or $\sim$1.68~$S_\oplus$), the silicate weathering
feedback may not be sufficient to offset CO$_2$ buildup at Super-Io
eruption rates. As the CO$_2$ partial pressure rises, the greenhouse
effect amplifies surface temperatures, potentially triggering a moist
greenhouse transition and initiating the positive water vapor feedback
that leads to complete atmospheric desiccation
\citep{kasting1988c,noack2017a}. Since the stellar irradiation of
GJ~12~b already places it at or above the runaway greenhouse threshold
under cloud-free, Earth-like albedo conditions
(Section~\ref{sec:energy_budget}), only a modest additional CO$_2$
forcing is required to cross the threshold even for a highly
reflective atmosphere. As described in Section~\ref{sec:prems}, a
period of high eccentricity and elevated stellar luminosity may have
already established super-volcanic outgassing rates early in the
system's history, depleting the mantle volatile inventory and
pre-conditioning the atmosphere for the runaway transition. The
volcanic outgassing pathway thus acts in concert with both the direct
thermal forcing and pre-MS scenarios, reinforcing the conclusion that
GJ~12~b most likely evolved into a Venus-like state.


\subsubsection{Dynamical Origins of Eccentricity}
\label{sec:dynamics}
 
Tidal circularization timescales for eccentric orbits with relatively
short orbital periods, such as GJ~12~b, are typically short compared
to the age of the system. The characteristic circularization timescale
for a rocky planet is \citep{jackson2008a}:
\begin{equation}
    \tau_{\rm circ} \approx \frac{4}{63} \frac{Q}{k_2}
    \frac{M_p}{M_\star} \left(\frac{a}{R_p}\right)^5
    \frac{P}{2\pi},
    \label{eq:circ_time}
\end{equation}
For GJ~12~b with $Q = 50$, $k_2 = 0.3$, $M_p = 0.95$~$M_\oplus$,
$M_\star = 0.255$~$M_\odot$, $R_p = 0.93$~$R_\oplus$: $\tau_{\rm circ}
\approx 9.7 \times 10^{9}$~yr. This is significantly longer than the
estimated stellar age of $\gtrsim$2.8~Gyr. The implication of this is
that, unlike many short-period planets, the eccentricity of GJ~12~b
has not necessarily been damped over the system's lifetime. If GJ~12~b
formed or was perturbed into an eccentric orbit, that eccentricity
could have persisted throughout the $\sim$2.8~Gyr history of the
system without requiring an ongoing excitation mechanism.

This result stands in sharp contrast to the dynamical situation of
bodies frequently invoked in the tidal heating literature. Io ($P =
1.77$~days), with its proximity to Jupiter and the Laplace resonance,
has a circularization timescale far shorter than the age of the Solar
System, and its eccentricity ($e \approx 0.004$) is maintained solely
through resonant forcing by Europa and Ganymede
\citep{peale1979,lainey2009}. Similarly, Titan ($P = 15.95$~days) has
a very modest eccentricity of 0.029 that likely requires resonant or
secular forcing to maintain. The ultra-short-period planet 55~Cnc~e
($P = 0.74$~days) has $\tau_{\rm circ} \ll t_{\rm age}$, meaning any
primordial eccentricity would have been damped long ago;
\citet{ferrazmello2025} showed that the observed nonzero eccentricity
of 55~Cnc~e is likely forced by an undetected companion in a close
orbital resonance. GJ~12~b, with its much longer orbital period of
12.76~days and correspondingly larger semi-major axis, occupies a
fundamentally different dynamical regime in which $\tau_{\rm circ}
\approx 9.7$~Gyr exceeds the system age. An ongoing forcing mechanism
is therefore not required, and the eccentricity is naturally
long-lived. This also means that the cumulative thermal evolution of
GJ~12~b under sustained tidal heating could span the entire history of
the system, as discussed further in Section~\ref{sec:prems}.

There are several mechanisms that could explain the current orbital
eccentricity of GJ~12~b. Given that $\tau_{\rm circ} \gg t_{\rm age}$,
the most straightforward interpretation is that the eccentricity is
either primordial or the result of a past gravitational perturbation,
such as a planet-planet scattering event or resonance passage. Both
scattering \citep{rasio1996c,chatterjee2008,ford2008c} and resonance
crossing \citep{chiang2002a} are well-established mechanisms for
imprinting eccentricities onto planetary orbits, and the resulting
eccentricity distribution of the observed exoplanet population is
broadly consistent with scattering models \citep{juric2008b}. In
either case, the eccentricity would be expected to persist to the
present day without requiring ongoing excitation, making the current
non-zero eccentricity entirely plausible from a dynamical standpoint.

A further explanation is eccentricity excitation via an undetected
companion planet through secular or resonant interactions. The current
RV datasets of \citet{brady2025b} and \citet{turner2026a} do not
exclude the presence of a Neptune-mass or sub-Saturn planet at orbital
periods of several tens of days. Secular forcing by an outer companion
is a well-known mechanism for sustaining non-zero eccentricities in
otherwise rapidly damping inner planets
\citep{correia2012a,laskar2012,kane2014b}, and statistical studies of
multi-planet systems predict that a substantial fraction of apparently
single planets host undetected companions \citep{dietrich2020}.
 

\subsubsection{Atmospheric Characterization with JWST}
\label{sec:jwst}
 
GJ~12~b has a transmission spectroscopy metric (TSM;
\citealt{kempton2018}) of $\sim$70, comparable to the TRAPPIST-1
planets, making it one of the best targets in the known terrestrial
planet population for JWST atmospheric characterization
\citep{dholakia2024,kuzuhara2024}. The combination of a low-activity
host star, proximity to Earth, and Earth-similar radius makes GJ~12~b
an exceptional test case for distinguishing between a Venus-like
post-runaway atmosphere and a temperate planet.
 
From the tidal Venus perspective, JWST observations of GJ~12~b can
test several hypotheses. If GJ~12~b underwent a runaway greenhouse
scenario, it would likely have developed a thick CO$_2$-dominated
atmosphere with negligible water vapor. JWST transmission spectroscopy
may show strong CO$_2$ absorption features at 4.3~$\mu$m and
potentially at 2.7 and 15~$\mu$m, with an absence of H$_2O$ features
\citep{lustigyaeger2019a,ostberg2023c}. However, if the runaway
greenhouse that desiccated GJ~12~b was followed by sustained pre-MS
EUV stripping of the resulting CO$_2$ atmosphere \citep{zahnle2017},
the planet might now lack any substantial atmosphere. In this case,
the thermal emission would closely track the stellar irradiation
pattern, with a large day-night temperature contrast. MIRI secondary
eclipse observations could distinguish this scenario from a Venus-like
atmosphere by the absence of spectral features and the high dayside
brightness temperature
\citep{morley2017b,koll2019b,greene2023,zieba2023}. Another scenario
is that of a temperate present-day atmosphere, requiring a high Bond
albedo ($A \gtrsim 0.5$) and a near-circular orbit to avoid tidal
heating exceeding the albedo-mediated threshold. JWST transmission
spectroscopy would show H$_2$O and CO$_2$ features consistent with a
water-bearing atmosphere \citep{morley2017b,mansfield2019}. Finally,
tidally induced volcanism could sustain a secondary atmosphere through
outgassing even if the primary atmosphere was stripped. Depending on
initial volatile inventories \citep{kane2020d,kite2020c}, the
resulting atmosphere may resemble a more volcanically active Venus
analog, with SO$_2$ absorption features at 7.3 and 8.7~$\mu$m
accessible to MIRI \citep{kaltenegger2010f,quick2020b,ostberg2023b}.

We estimate that 5--10 JWST transits with NIRSpec PRISM
(0.6--5.3~$\mu$m) would be sufficient to detect or rule out a
CO$_2$-dominated Venus-like atmosphere for GJ~12~b, based on scaling
relations from \citet{lustigyaeger2019a} and
\citet{turbet2023}. Secondary eclipse observations with MIRI
(5--28~$\mu$m) would further constrain the atmospheric composition and
thermal structure. A caveat remains that the ``cosmic shoreline''
framework \citep{zahnle2017} predicts that planets with GJ~12~b's low
mass (0.95~$M_\oplus$) and high stellar EUV flux exposure may struggle
to retain even secondary atmospheres. However, the low current
activity of GJ~12 \citep{kuzuhara2024} suggests that any residual
atmosphere may be relatively stable at the present epoch. JWST
observations will provide a critical test of whether low-mass
terrestrial planets around M dwarfs can retain atmospheres under such
conditions.

The Rocky Worlds Director's Discretionary Time (DDT) program, which
devotes 500~hours of JWST/MIRI time to secondary eclipse observations
of $\sim$12 rocky M-dwarf planets at 15~$\mu$m, will provide
systematic constraints on the atmospheric survival rate for this class
of planets. The DDT targets were selected using the cosmic shoreline
as a priority metric, and several planets in the program share key
properties with GJ~12~b. Results from the first DDT target, GJ~3929~b,
indicate a bare-rock surface, consistent with the expectation that
highly irradiated rocky M-dwarf planets experience significant
atmospheric loss. GJ~12~b's lower equilibrium temperature and
quiescent host star make it a comparatively favorable case for
atmospheric retention, and it represents an important complement to
the DDT sample. Recent work by \citet{zilinskas2025} on 55~Cnc~e has
demonstrated how JWST emission spectroscopy can constrain atmospheric
composition and volatile outgassing on tidally heated rocky planets,
providing a methodological template for characterizing GJ~12~b's
atmospheric state.


\subsection{Comparative Observational Prospects}
\label{sec:obs}

The case studies presented above represent distinct categories in
the tidal Venus taxonomy, and observational follow-up can distinguish
their predicted outcomes.

For TOI-6716~b and TOI-912~b, the immediate priority is RV follow-up
to confirm or refine the orbital eccentricities that determine whether
these planets are genuinely tidally dominated. If the eccentricities
are confirmed, thermal emission measurements with JWST/MIRI could
probe the extreme surface temperatures expected for tidally dominated
worlds. If the eccentricities are revised downward, these planets
would transition to the flux-driven Venus analog category, illustrating
the sensitivity of the tidal Venus classification to orbital parameter
precision.

For TOI-700~d, the critical question is whether a temperate atmosphere
survived the $\sim$0.5--1~Gyr pre-MS above-threshold epoch. JWST
transmission spectroscopy with NIRSpec can test for H$_2$O and CO$_2$
features consistent with a water-bearing atmosphere
\citep{morley2017b,lustigyaeger2019a}. Detection of a temperate
atmosphere would demonstrate that Earth-mass planets at the inner HZ
boundary of M2 dwarfs can survive the pre-MS gauntlet, providing an
important constraint on the Category~3 (historically compromised)
classification. The quiescent nature of the host star, with no
detected flares, makes TOI-700~d among the most favorable cases for
atmospheric retention \citep{gilbert2020b}.

For LHS~1140~b, JWST programs are already underway and initial results
suggest possible atmospheric features \citep{cadieux2024a}. As a
5.6~$M_\oplus$ super-Earth with high surface gravity, LHS~1140~b
provides the most stringent test of whether high-mass terrestrial
planets retain atmospheres through multi-Gyr pre-MS irradiation. The
predicted outcomes differ sharply depending on the atmospheric state:
a water-world or ice-world scenario
\citep{damiano2024a,cadieux2024a} would indicate successful volatile
retention despite above-threshold pre-MS irradiation, while a bare
rock would suggest that even super-Earth masses are insufficient for
atmospheric survival around late M dwarfs.

For GJ~12~b, the observational diagnostics focus on distinguishing
among the four primary outcomes described in Section~\ref{sec:jwst}:
a CO$_2$-dominated Venus-like atmosphere, a volcanically sustained
SO$_2$/CO$_2$ secondary atmosphere, a bare-rock surface, or
(least likely) a temperate water-bearing atmosphere. The 3D climate
simulations (Section~\ref{sec:gcm}) provide quantitative predictions
that can be directly tested.

Together, these five systems provide a set of complementary tests of
the tidal Venus framework: confirmation of the eccentricities of
TOI-6716~b and TOI-912~b would establish the reality of the tidally
dominated regime; atmospheric detections on TOI-700~d and LHS~1140~b
would constrain whether planets survive pre-MS irradiation; and
GJ~12~b tests the atmospheric signatures of the tidally compounded
Venus state. The Rocky Worlds DDT program will provide additional
context by constraining the bare-rock fraction among a larger sample
of rocky M-dwarf planets.


\section{Discussion}
\label{sec:disc}


\subsection{How Many Tidal Venuses Exist?}
 
From Section~\ref{sec:dist}, of the 143 terrestrial exoplanets ($R_p <
2\,R_\oplus$) with measured orbital eccentricities in our sample, 137
(96\%) already exceed the runaway greenhouse limit from their total
energy budget. This result is driven primarily by the strong
observational bias toward short-period planets in the current
exoplanet census \citep{fressin2013,dressing2015b}. The more
physically interesting subset consists of planets that are pushed past
a threshold by tidal heating that they would not otherwise cross from
stellar flux alone. Eight planets (5.6\%) exceed the molten surface
limit of $\sim$370~kW/m$^2$, with K2-211~b, K2-147~b, TOI-1238~b, and
GJ~367~b achieving tidal fluxes exceeding $10^7$~W/m$^2$, several
orders of magnitude above the extreme volcanism threshold and
representative of the most extreme tidal environments in the known
exoplanet census \citep{goffo2023}. A total of 100 planets (70\%)
exceed the extreme volcanism threshold of 2~W/m$^2$ in tidal flux
alone, highlighting that Super-Io geological activity may be common
among short-period rocky worlds \citep{driscoll2015,noack2017a}, even
for many planets whose total flux already places them well above the
runaway greenhouse limit.

The most important tidal Venus candidates for habitability
considerations are the six planets that fall below the runaway
greenhouse limit (Kepler-441~b, Kepler-186~f
\citep{quintana2014a}, LHS~1140~b \citep{dittmann2017a},
Kepler-296~f, Kepler-442~b, and TOI-700~d \citep{gilbert2020b})
since these are potentially habitable worlds where eccentricity-driven
tidal heating could, in principle, render them uninhabitable. Of
these, all have tidal fluxes many orders of magnitude below the
extreme volcanism threshold, so they are not currently endangered by
either the direct or volcanic outgassing pathways. As demonstrated
for TOI-700~d and LHS~1140~b in Section~\ref{sec:hist_comp}, the
$a^{-15/2}$ distance scaling of tidal dissipation means that no
physical eccentricity can bridge the gap to the runaway greenhouse
threshold at their current orbital distances. However, this
reflects only their present dynamical state, and all six qualify as
Category~3 (historically compromised) planets in the taxonomy of
Section~\ref{sec:taxonomy}: if any of these planets
harbored higher eccentricities in the past due to dynamical
interactions with unseen companions, the combination of elevated
pre-MS stellar irradiation and tidal heating may have crossed the
runaway greenhouse threshold during an earlier epoch, permanently
desiccating them, a reminder that present-day habitability assessments
based on insolation alone are insufficient
\citep{kane2014e,ostberg2023a}.

An important implication is that the fraction of habitable worlds
among temperate M dwarf planets may be substantially lower than
previous estimates based on insolation alone
\citep{dressing2015b}. The TRAPPIST-1 planets
\citep{gillon2017a,agol2021}, which orbit in a compact,
near-resonant configuration, have nearly circular orbits due to
mutual tidal damping and may not suffer from the tidal Venus
phenomenon, despite their close orbital separations
\citep{barr2018}. However, planets in systems with more dynamically
active architectures, particularly those with giant planet companions
in eccentric orbits, are at much higher risk, and the current sample
of 143 planets demonstrates that extreme tidal environments are not
rare among the confirmed population.


\subsection{The Role of Pre-Main-Sequence Activity}
 
As discussed in the context of our case studies
(Sections~\ref{sec:hist_comp} and \ref{sec:prems}), the
pre-MS active phase of M dwarf host stars represents a critical and
often underappreciated threat to terrestrial planet habitability. M
dwarfs are substantially more luminous in the X-ray and
extreme-ultraviolet (XUV) during the first $\sim$1--3~Gyr of their
lives, with XUV fluxes orders of magnitude above present-day values
\citep{ribas2005,france2016a,wheatley2017}. This sustained
high-energy irradiation drives rapid atmospheric escape and can
desiccate planets that would otherwise be considered temperate at
current stellar luminosities \citep{ramirez2014c,luger2015b}. For the
broader population of tidal Venus candidates, this consideration
amplifies the conclusions of the demographic analysis. Planets that
are currently near the runaway greenhouse threshold from stellar
irradiation alone were in all likelihood exposed to considerably more
extreme conditions during the pre-MS epoch. If those planets also
harbored non-zero eccentricities at early times, the historical tidal
contribution may have been decisive. For planets with long
circularization timescales (as in the case of GJ~12~b) such
eccentricities persist naturally without requiring ongoing excitation,
meaning tidal heating could have operated continuously throughout the
entire pre-MS and main-sequence history. In this sense, the tidal
Venus mechanism is not only a present-day process but also a
historical one, in that many planets classified as potentially
habitable today may have had their fates sealed during the pre-MS
epoch, from which their water inventories could not recover. The
three-category taxonomy introduced in Section~\ref{sec:taxonomy}
captures this temporal dimension: the distinction between tidally
dominated, historically compromised, and tidally compounded planets
reflects not only the present-day energy budget but also the
evolutionary history that determines whether a planet ever had the
opportunity to develop and retain a habitable environment.


\subsection{Eccentric Terrestrial Planets in Exoplanetary Systems}
 
GJ~12~b's potential eccentricity raises broader questions about the
prevalence of eccentric terrestrial planets in other planetary
systems. RV surveys have established that a substantial fraction of
giant planets reside on eccentric orbits
\citep{butler2006,cumming2008,winn2015}, and dynamical modeling shows
that such planets can excite and maintain non-zero eccentricities in
interior terrestrial planets through secular interactions
\citep{barnes2013a,kane2024c}. Even in the absence of a giant planet
companion, scattering events during the late stages of planet
formation can imprint eccentricities that persist for gigayears in
systems with long circularization timescales
\citep{rasio1996c,chatterjee2008}. The LP~791-18 system provides a
direct observational example of secular architecture driving tidal
activity, with the outer sub-Neptune forcing the inner temperate
Earth-sized planet onto an eccentric orbit with significant tidal
heating \citep{peterson2023}.
 
Future RV surveys with instruments such as ESPRESSO, NEID, and
MAROON-X will likely identify additional planets in the GJ~12
system. If a companion planet is found, particularly at an orbital
period of 20--100 days, then this would provide an additional
dynamical context for GJ~12~b's eccentricity, although the long
circularization timescale means a companion is not required to sustain
the eccentricity over the system's lifetime. GJ~12~b already
represents a compelling archetype for the tidal Venus phenomenon since
it is an Earth-sized planet whose naturally persistent eccentricity,
combined with the elevated pre-MS stellar luminosity, plausibly drove
complete atmospheric desiccation early in the system's history.

 
\subsection{Connections to Venus's History}

The GJ~12 system provides a compelling exoplanetary analog for the
kind of scenario that may have driven Venus's climate catastrophe. The
geological and atmospheric record of Venus strongly implies that it
once harbored liquid water, as evidenced by the high deuterium-to-hydrogen
ratio in its atmosphere \citep{donahue1982,donahue1999}. The
mechanism responsible for Venus's transition from a potentially
habitable world to its present state, a 735~K surface beneath a
92-bar CO$_2$ atmosphere, remains debated. Leading hypotheses invoke
the combined effects of a runaway or moist greenhouse triggered by
elevated solar luminosity over geological time
\citep{hamano2013,turbet2021}, volcanic resurfacing that terminated
the carbon-silicate cycle \citep{way2020,gillmann2020}, and late
volatile delivery that modulated the atmospheric inventory
\citep{gillmann2020}. If Jupiter migrated through the inner solar
system early in its history and excited Venus to an eccentricity of
$e \sim 0.3$ \citep{kane2020e}, tidal heating would have provided
an additional energy source contributing to rapid water loss. The key
question is whether Venus's current near-circular orbit is the
endpoint of tidal circularization following a high-eccentricity phase,
or whether Earth itself played a role in damping Venus's eccentricity
through secular interactions \citep{laskar2012}.
 
For GJ~12~b, the analogous question is whether the recovered
eccentricity of $0.24 \pm 0.11$ \citep{brady2025b,turner2026a} is
primordial or the result of a past dynamical perturbation. Given the
long circularization timescale of $\sim$9.7~Gyr, the eccentricity is
naturally long-lived regardless of origin, and GJ~12~b could have
been actively tidally heated for most or all of the system's history.
Whether the eccentricity is real or consistent with zero within the
measurement uncertainties, the pre-MS argument ensures that the planet
was exposed to extreme combined flux during the first $\sim$1--3~Gyr,
with JWST poised to reveal the atmospheric outcome.

Understanding the history of Venus is therefore directly relevant to
interpreting the nature of potential Venus analogs throughout the
galaxy. Venus serves as the only ground-truth example of a runaway
greenhouse outcome for a terrestrial planet of known size, mass, and
orbital location, making it an essential calibration point for
exoplanet atmospheric models \citep{kane2019d,jakosky2025b}. The
demographics of VZ planets around other stars have grown substantially
with TESS \citep{ostberg2019,ostberg2023a}, but without an empirical
understanding of how Venus underwent climate divergence, it is
difficult to reliably predict which of those exoplanets share Venus's
fate.

There are several planned upcoming missions to Venus, including the
NASA VERITAS (Venus Emissivity, Radio Science, InSAR, Topography, and
Spectroscopy) \citep{cascioli2021} and DAVINCI (Deep Atmosphere Venus
Investigation of Noble gases, Chemistry, and Imaging)
\citep{garvin2022} missions, and the ESA EnVision spacecraft
\citep{widemann2023}. Together, these missions will provide
transformative new constraints on Venus's geologic timeline,
atmospheric composition, surface mineralogy, and interior structure.
DAVINCI's descent probe will directly sample noble gas abundances and
isotopic ratios that encode the history of Venus's volatile evolution,
including whether Venus ever had a surface ocean and when it lost its
water \citep{garvin2022}. VERITAS will map the surface at
unprecedented resolution to search for evidence of recent volcanism,
active tectonics, and the timescales of resurfacing
\citep{cascioli2021}. EnVision will characterize the structure and
dynamics of Venus's atmosphere and probe surface–atmosphere exchange
processes \citep{widemann2023}. The data from these three missions
will directly inform the study of volcanic outgassing rates and
atmospheric escape mechanisms that are central to the tidal Venus
mechanism. In particular, they will test whether the volcanic
outgassing pathway described in Section~\ref{sec:volcanic} has an
analog in Venus's own geological history. The degree to which Venus's
evolution was governed by internal heat sources versus external
irradiation will inform how we weight tidal versus stellar
contributions in the energy budgets of exoplanet Venus analogs
\citep{gillmann2022}. GJ~12~b, as an Earth-mass planet in a Venus-like
irradiation environment with a measurable eccentricity, represents an
important system that such ground-truth calibration is needed to
interpret.


\section{Conclusions}
\label{sec:con}

The tidal evolution of rocky exoplanets represents an underexplored
but consequential pathway for transforming potentially habitable worlds
into Venus analogs. This paper investigates the demographic
landscape of tidal Venus candidates among the known terrestrial
exoplanet population and develops a three-threshold framework
governing the tidal Venus phenomenon: the extreme volcanism limit
($F_{\rm tide} \geq 2$~W/m$^2$), the runaway greenhouse limit
($F_{\rm total} \geq 295$~W/m$^2$; \citealt{heller2015b}), and the
molten surface limit ($F_{\rm total} \geq 3.7\times10^5$~W/m$^2$).
An analysis of 143 terrestrial exoplanets with measured eccentricities
reveals that 96\% already exceed the runaway greenhouse limit, 70\%
exceed the extreme volcanism threshold in tidal flux alone, and 8
systems exceed the molten surface limit. The six planets that fall
below the runaway greenhouse threshold (Kepler-441~b, Kepler-186~f,
LHS~1140~b, Kepler-296~f, Kepler-442~b, and TOI-700~d) represent the
subset of known temperate worlds for which the tidal Venus mechanism
poses the greatest prospective threat, should their eccentricities
have been elevated during any prior dynamical episode.

We introduce a three-category taxonomy for tidal influence on
planetary climate: (1) flux-driven Venus analogs, where stellar
irradiation alone exceeds the runaway greenhouse threshold and tidal
heating compounds the Venus-like state through volcanism and
outgassing; (2) tidally dominated planets, where tidal flux alone
exceeds the runaway greenhouse threshold; and (3) historically
compromised HZ planets, which were exposed to above-threshold
conditions during the host star's pre-MS phase. Five case studies
illustrate this taxonomy.

TOI-6716~b and TOI-912~b exemplify the tidally dominated regime, with
tidal fluxes of $\sim$17,400 and $\sim$42,100~W/m$^2$ that exceed
their stellar fluxes by factors of $\sim$10 and $\sim$8,
respectively. These extreme tidal energy budgets depend on orbital
eccentricities that require confirmation through further RV follow-up,
but if verified, these planets represent the clearest examples of the
tidal Venus phenomenon. TOI-700~d is the most marginal HZ planet in
our sample, sitting $\sim$3~W/m$^2$ below the runaway greenhouse
threshold. While tidal heating at its orbital distance (0.161~AU) is
too weak to bridge this gap at any physical eccentricity, the
$\sim$0.5--1~Gyr pre-MS phase of its M2 host star subjected the planet
to stellar fluxes $\sim$3--10$\times$ above the current value, well in
excess of the runaway threshold. Whether TOI-700~d retained its
atmosphere through this epoch is the central question for its
habitability. LHS~1140~b, a 5.6~$M_\oplus$ super-Earth deep in the
conservative HZ, was likewise exposed to above-threshold irradiation
during a $\sim$1--3~Gyr pre-MS epoch. Its high mass likely enables
superior atmospheric retention, making it the most stringent test of
whether massive terrestrial planets survive the M~dwarf pre-MS
gauntlet. Further JWST observations will determine whether an
atmosphere persists.

GJ~12~b is an excellent example of a flux-driven Venus analog with
significant tidal compounding. Its stellar insolation of
$\sim$1.68~$S_\oplus$ already exceeds the runaway greenhouse
threshold, while its tidal flux of $\sim$5.4~W/m$^2$ drives Super-Io
volcanism and an independent geochemical pathway to a runaway
greenhouse. Three-dimensional climate simulations with the Generic-PCM
show that a temperate atmosphere fails to achieve radiative balance,
while a Venus-like CO$_2$-dominated atmosphere converges to a stable
state at 430~K. The tidal circularization timescale of $\sim$9.7~Gyr,
far exceeding the system age, means GJ~12~b's eccentricity is
naturally long-lived and tidal heating may have operated continuously
throughout the system's history.

Together, these five systems provide a complementary set of testable
predictions: RV follow-up of TOI-6716~b and TOI-912~b will establish
whether the tidally dominated regime is real or an artifact of
eccentricity upper limits; JWST observations of TOI-700~d and
LHS~1140~b test whether planets survive pre-MS irradiation; and
GJ~12~b tests the atmospheric signatures of the tidally compounded
Venus state. The forthcoming
DAVINCI, VERITAS, and EnVision missions to Venus
\citep{garvin2022,cascioli2021,widemann2023} will provide
ground-truth constraints on volcanic outgassing, atmospheric escape,
and the runaway greenhouse, directly calibrating the models needed to
interpret JWST observations of these exoplanet Venus candidates. As
the inventory of well-characterized temperate terrestrial exoplanets
grows through TESS and JWST, the tidal Venus framework developed here
provides a systematic tool for assessing which worlds' fates may have
been sealed by the combined action of tidal heating and elevated
pre-main-sequence irradiation during the earliest epochs of their
systems' histories.


\section*{Acknowledgements}

The authors would like to thank the anonymous referee, whose feedback
helped to improve the manuscript. This research has made use of the
NASA Exoplanet Archive, which is operated by the California Institute
of Technology, under contract with the National Aeronautics and Space
Administration under the Exoplanet Exploration Program. This research
has also made use of the Habitable Zone Gallery at hzgallery.org. The
results reported herein benefited from collaborations and/or
information exchange within NASA's Nexus for Exoplanet System Science
(NExSS) research coordination network sponsored by NASA's Science
Mission Directorate.


\software{Generic-PCM \citep{wordsworth2010b,wordsworth2011}}




\end{document}